\documentclass[12pt]{article}
\usepackage{graphicx,amssymb,amsmath}

\newcommand{\be}{\begin{equation}}
\newcommand{\ee}{\end{equation}}
\newcommand{\bea}{\begin{eqnarray}}
\newcommand{\eea}{\end{eqnarray}}
\newcommand{\beas}{\begin{eqnarray*}}
\newcommand{\eeas}{\end{eqnarray*}}

\newcommand{\xm}{x^-}
\newcommand{\xmp}{x^-{}'}
\newcommand{\xp}{x^+}
\newcommand{\xpp}{x^+{}'}

\begin{document}
\begin{titlepage}

\vspace*{-24mm}

\rightline{NSF-KITP-12-045}

\vspace{4mm}

\begin{center}

{\Large Holographic representation of bulk fields with spin in AdS/CFT}

\vspace{4mm}

\renewcommand\thefootnote{\mbox{$\fnsymbol{footnote}$}}
Daniel Kabat${}^{1}$\footnote{daniel.kabat@lehman.cuny.edu},
Gilad Lifschytz${}^{2,3}$\footnote{giladl@research.haifa.ac.il},
Shubho Roy${}^{1,4}$\footnote{sroy@ccny.cuny.edu} and
Debajyoti Sarkar${}^{1,5}$\footnote{dsarkar@gc.cuny.edu}

\vspace{4mm}

${}^1${\small \sl Department of Physics and Astronomy} \\
{\small \sl Lehman College of the CUNY, Bronx NY 10468, USA}

${}^2${\small \sl Department of Mathematics and Physics} \\
{\small \sl University of Haifa at Oranim, Kiryat Tivon 36006, Israel}

${}^3${\small \sl Kavli Institute for Theoretical Physics} \\
{\small \sl University of California, Santa Barbara, CA 93106-4030}

${}^4${\small \sl Physics Department} \\
{\small \sl City College of the CUNY, New York NY 10031, USA}

${}^5${\small \sl Graduate School and University Center} \\
{\small \sl City University of New York, New York NY 10036, USA}

\end{center}

\vspace{4mm}

\noindent
We develop the representation of bulk fields with spin one and spin two in anti-de Sitter space,
as non-local observables in the dual CFT.  Working in holographic gauge in the bulk, at leading order
in $1/N$ bulk gauge fields are obtained by smearing boundary
currents over a sphere on the complexified boundary, while linearized metric fluctuations are obtained by
smearing the boundary stress tensor over a ball.  This representation respects AdS covariance up to a
compensating gauge transformation.  We also consider massive vector fields, where the bulk field is
obtained by smearing a non-conserved current.  We compute bulk two-point functions
and show that bulk locality is respected.  We show how to include interactions of massive vectors using
$1/N$ perturbation theory, and we comment on the issue of general backgrounds.

\end{titlepage}
\setcounter{footnote}{0}
\renewcommand\thefootnote{\mbox{\arabic{footnote}}}

\section{Introduction\label{sect:intro}}

The question of locality and causality in quantum gravity is an old and unresolved issue.
AdS/CFT implies that at best locality and causality are approximate notions. However it is vital to understand in what situations and in what way the notion of bulk locality arises.
One approach to this issue, pursued since the early days of AdS/CFT, is to construct operators in the CFT which can mimic
the local field operators of bulk supergravity.

In \cite{Balasubramanian:1998sn,Banks:1998dd,Dobrev:1998md,Bena:1999jv} free scalar fields in the bulk were expressed as CFT operators,
and it was shown that bulk locality was obeyed in the leading large-$N$ limit.
This approach was refined to obtain CFT expressions that are covariant and convenient in \cite{Hamilton:2005ju,Hamilton:2006az,Hamilton:2006fh}. In particular it was shown that one can represent bulk scalar fields as smeared operators in the CFT, where the smearing has support on a ball on the complexified boundary.
In \cite{Kabat:2011rz} it was shown that for scalar fields this construction can be extended include interactions using $1/N$ perturbation theory.  
The construction of bulk operators in asymptotically AdS spacetimes has been further extended and clarified in \cite{Heemskerk:2012mn}.

In this paper we build upon two approaches that have been successfully used to construct scalar fields in the bulk.
\begin{enumerate}
\item
Given a bulk Lagrangian one can solve the bulk equations of motion perturbatively, to express the Heisenberg picture field operators in
terms of boundary data.  This leads to an expression for the bulk field as a sum of smeared CFT operators.
The bulk operator constructed in this way of course respects locality, assuming one starts from a local Lagrangian in the bulk,
but the construction seems tied to knowing the bulk
equations of motion.
\item
Alternatively one can start in the CFT with a candidate bulk operator, constructed by solving free equations of motion, then demand that bulk micro-causality holds at the level of three point functions. This can be achieved order-by-order in the $1/N$ expansion, by modifying the
definition of the bulk field in the CFT to include a sum of appropriately-smeared higher dimension operators.  In this construction the guiding principle is bulk micro-causality.
\end{enumerate}
The later construction can be carried out fully inside the CFT, without knowing the bulk Lagrangian. Hence it may enable one to see the limitations of bulk perturbation theory, and
understand the way in which micro-causality breaks down at the non-perturbative level.  A difficulty of extending the
second approach to gauge fields is that the correct statement of bulk micro-causality is necessarily somewhat subtle \cite{Heemskerk:2012mn}.

An outline of this paper is as follows.
In the first part of this paper we extend the program of \cite{Hamilton:2005ju,Hamilton:2006az,Hamilton:2006fh} to free fields with spin one and spin two.
A closely related construction has been carried out by Heemskerk \cite{Heemskerk:2012mq}.
In section \ref{sect:gaugesmear} we derive the smearing function for a bulk gauge field and show that it is
covariant under conformal transformations. We compute the bulk-to-boundary two point function and show that, although the gauge field does not obey micro-causality, the corresponding field strength does.
In section \ref{sect:metric} we obtain analogous results for gravity: we work out the smearing function for a graviton and show that
the graviton has non-local correlators.  In the context of gravity it is the Weyl tensor that obeys bulk micro-causality.
In section \ref{sect:massivevector} we derive the smearing function for a massive vector field, and show that a massive vector directly
obeys micro-causality.  The helps clarify the relation between gauge symmetry and locality.

In the second part of this paper we discuss interactions and general backgrounds.
In section \ref{sect:interaction} we show how to extend the definition of a massive vector field in the bulk to include interactions,
using perturbation theory in $1/N$, and we discuss the difficulty with gauge fields resulting from the existence of conserved charges.
In section \ref{sect:generalbackground} we provide a framework for extending the construction to general backgrounds and for going
beyond the approximation of having a fixed background.  We also explain the necessary conditions for the existence of approximately local operators in the bulk.

\section{Gauge smearing functions\label{sect:gaugesmear}}

In this section we develop the representation of an abelian bulk gauge field
as a non-local observable in the dual CFT.  Our basic result is given in (\ref{gaugesmear})
below: the bulk gauge field at a point $(x,z)$ in the bulk is obtained by integrating the
boundary current over a sphere of radius $z$ on the complexified boundary.

Our conventions are as follows.  We work in Poincar\'e coordinates in AdS${}_{d+1}$ with metric
\[
ds^2 = G_{MN} dX^M dX^N = {R^2 \over z^2}\left(\eta_{\mu\nu} dx^\mu dx^\nu + dz^2\right) \qquad \mu,\nu = 0,\ldots,d-1
\]
The boundary at $z = 0$ carries a flat Minkowski metric, $\eta_{\mu\nu} = {\rm diag}(-+\cdots+)$.  Boundary
indices $\mu,\nu$ are raised and lowered with $\eta_{\mu\nu}$.

Our goal is to solve the source-free Maxwell equations in the bulk, $\nabla_M F^{MN} = 0$,
with the boundary conditions
\be
\label{Fbc}
F_{z \mu}(x,z) \sim (d-2) z^{d-3} j_\mu(x) \qquad \hbox{\rm as $z \rightarrow 0$} 
\ee
The factor $d-2$ is inserted for later convenience.\footnote{The special case $d = 2$ will be discussed in section \ref{sect:CS}.}
From the bulk perspective this defines $j_\mu(x)$ as the coefficient of the leading small-$z$
behavior of the bulk field.  But in the dual CFT $j_\mu(x)$ is interpreted as a conserved current.
So if we can solve for the bulk field in terms of its near-boundary behavior, via a kernel of the form
\be
\label{gaugesmear1}
A_M(x,z) = \int d^dx' \, K_M{}^\mu(x,z \vert x') j_\mu(x')\,,
\ee
then we will have succeeded in representing the bulk gauge field as a non-local observable in the dual CFT.
We'll refer to $K_M{}^\mu$ as a smearing function, although as we'll see below, smearing distribution might
be more appropriate.

A few comments are in order.
\begin{itemize}
\item
The smearing function we are after should not be confused with Witten's bulk-to-boundary propagator,
which relates a non-normalizeable field in the bulk to a source in the dual CFT \cite{Witten:1998qj}.  Rather we
wish to express a {\em normalizeable} field in the bulk in terms of an {\em operator} in the CFT.
\item
The AdS boundary is timelike, so this is not a standard Cauchy problem.  Nonetheless, in all cases of
interest, it seems an explicit solution is possible.  There is some discussion of this fact in \cite{Heemskerk:2012mn}.  Also note that
we will construct smearing functions with compact
support on the complexified boundary, along the lines of \cite{Hamilton:2006fh}.  For a construction with support on a real section of the boundary
see \cite{Heemskerk:2012mq}.
\end{itemize}

Of course the CFT doesn't know about bulk gauge symmetries -- it only sees global conservation laws -- so
in order to reconstruct a bulk gauge field we will need to make some choice of gauge in the bulk.  It's
convenient to work in ``holographic gauge'' and set
\[
A_z(x,z) = 0\,.
\]
This allows a residual gauge freedom
\[
A_\mu(x,z) \rightarrow A_\mu(x,z) + \partial_\mu \lambda(x)
\]
where the gauge parameter $\lambda$ is independent of $z$.  The equation of motion from varying $A_z$ is
\[
\partial_z\left(\eta^{\mu\nu} \partial_\mu A_\nu\right) = 0\,.
\]
Thus $\partial_\mu A^\mu$ is independent of $z$, and we can use a residual gauge transformation to set
$\partial_\mu A^\mu = 0$ everywhere.\footnote{From the CFT point of view this is guaranteed by the boundary conditions at $z = 0$,
where the bulk gauge field approaches a conserved current in the CFT.}  The remaining Maxwell equations then simplify to
\[
\partial_\mu \partial^\mu A_\nu + z^{d-3} \partial_z {1 \over z^{d-3}} \partial_z A_\nu = 0\,.
\]
Defining $\phi_\mu(x,z) = z A_\mu(x,z)$ one finds that\footnote{This amounts to expressing the
gauge field in a vielbein basis, setting $A_a = e_a{}^\mu A_\mu$ where $e_a{}^\mu = {z \over R} \delta_a{}^\mu$.}
\be
\label{ScalarWave}
\partial_\mu \partial^\mu \phi_\nu + z^{d-1} \partial_z {1 \over z^{d-1}} \partial_z \phi_\nu + {d - 1 \over z^2} \phi_\nu = 0\,.
\ee
This shows that each component of $\phi$ obeys the usual scalar wave equation,\footnote{The mass term actually represents a non-minimal coupling to curvature, $\left(\Box + \xi R\right)\phi = 0$ where $\xi = - {d-1 \over d(d+1)}$.} and from the mass term we can read off $m^2 R^2 = 1 - d$.

Although tachyonic, the scalar satisfies the BF bound \cite{Breitenlohner:1982jf}.  It is dual to an operator of conformal dimension
\[
\Delta = {d \over 2} + \sqrt{{d^2 \over 4} + m^2 R^2} = d - 1
\]
The normalizeable near-boundary behavior for such a scalar field is
\[
\phi_\mu(x,z) \sim z^{d-1} j_\mu(x) \qquad \hbox{\rm as $z \rightarrow 0$}
\]
In appendix \ref{sect:scalar} we show how to construct a smearing function for
such a scalar field.  The result, given in (\ref{scalarsmear}), can be used to represent a bulk gauge field in terms
of the boundary current.
\bea
\label{gaugesmear}
&& z A_\mu(t,{\bf x},z) =  {1 \over {\rm vol}(S^{d-1})} \hspace{-7mm}
\int\limits_{\hspace{8mm}t'{}^2 + \vert {\bf y}'\vert^2 = z^2} \hspace{-5mm} dt' d^{d-1} y' \,
j_\mu(t + t', {\bf x} + i {\bf y}') \\
\nonumber
&& {\rm vol}(S^{d-1}) = {2 \pi^{d/2} \over \Gamma(d/2)}
\eea
Here we're splitting the boundary coordinates $x^\mu = (t;{\bf x})$ into a time coordinate $t$
and $d-1$ spatial coordinates ${\bf x}$.  Note that the boundary current is evaluated
at complex values of the spatial coordinates.  The integral is over a sphere of
radius $z$ on the complexified boundary, with the center of the sphere located at $(t;{\bf x})$.

The basic claim is that (\ref{gaugesmear}) gives a gauge field that satisfies Maxwell's equations
and has the boundary behavior
\be
\label{Abc}
A_\mu(x,z) \sim z^{d-2} j_\mu(x) \qquad \hbox{\rm as $z \rightarrow 0$}
\ee
The fact that $A_\mu$ satisfies Maxwell's equations follows from appendix \ref{sect:scalar}, while
the boundary conditions are easy to check.  As $z \rightarrow 0$ the integration region shrinks to a point, so
we can bring the current outside the integral; the factors of ${\rm vol}(S^{d-1})$ cancel and we're left with
(\ref{Abc}).  The corresponding field strength then satisfies (\ref{Fbc}).  This is one nice feature of working on
the complexified boundary: it's manifest that local fields in the bulk go over to local operators in the CFT, in the
limit that the bulk point approaches the boundary.

Finally note that (\ref{gaugesmear}) can be written in a covariant form.  The invariant distance between
two points in AdS is
\[
\sigma(x,z \vert x',z') = {z^2 + z'{}^2 + (x - x')_\mu (x - x')^\mu \over 2 z z'}\,.
\]
The invariant distance diverges as $z' \rightarrow 0$.  However we can define a regulated bulk - boundary distance
\be
\label{sigmaz'}
(\sigma z')_{z' \rightarrow 0} = {z^2 + (x - x')_\mu (x - x')^\mu \over 2 z}
\ee
In terms of $\sigma z'$, the smearing integral (\ref{gaugesmear}) can be written as
\be
\label{gaugesmear2}
z A_\mu(t,{\bf x},z) = {1 \over {\rm vol}(S^{d-1})} \int dt' d^{d-1} y' \, \delta(\sigma z') \, j_\mu(t + t', {\bf x} + i {\bf y}')
\ee

\subsection{AdS covariance for gauge fields\label{sect:gaugecovariance}}

It's instructive to check that the smearing function (\ref{gaugesmear2}) behaves covariantly under conformal transformations.
First note that it's manifestly covariant under Poincar\'e transformations of the $x^\mu$ coordinates.  Under a dilation, which
corresponds to the bulk isometry
\[
x^\mu \rightarrow x'{}^\mu = \lambda x^\mu \qquad\quad z \rightarrow z' = \lambda z
\]
we have
\[
A_\mu \rightarrow A_\mu' = {1 \over \lambda} A_\mu \qquad\quad A_z \rightarrow A_z' = {1 \over \lambda} A_z
\]
Thus holographic gauge is preserved, $A_z' = 0$, and the quantity $zA_\mu$ appearing on the left hand side of (\ref{gaugesmear2})
transforms like a scalar.  This is consistent with the right hand side of (\ref{gaugesmear2}), since under a dilation $d^dx$ has dimension $-d$,
$\delta(\sigma z')$ has dimension $1$, and $j_\mu$ has dimension $d-1$.

Special conformal transformations are a little more subtle.  These correspond to the bulk isometry
\bea
&& x^\mu \rightarrow x'{}^\mu = {x^\mu - b^\mu (x^2 + z^2) \over 1 - 2 b \cdot x + b^2 (x^2 + z^2)} \\
&& z \rightarrow z' = {z \over 1 - 2 b \cdot x + b^2 (x^2 + z^2)}
\eea
Starting from holographic gauge $A_z = 0$ and working to first order in $b^\mu$ we find
\bea
&& A_z' = 2 z b \cdot A \\
\label{Amuprime}
&& A_\mu' = A_\mu + 2 x_\mu b \cdot A - 2 b_\mu x \cdot A - 2 b \cdot x A_\mu
\eea
So holographic gauge isn't preserved.  To restore it we make a compensating gauge transformation $A \rightarrow A + d \lambda$ where
\[
\lambda = - {1 \over {\rm vol}(S^{d-1})} \int d^dx' \, \theta(\sigma z') \, 2 b \cdot j
\]
The gauge parameter $\lambda$ has been chosen so that
\be
\partial_z \lambda = - {1 \over {\rm vol}(S^{d-1})} \int d^dx' \, \delta(\sigma z') \, 2 b \cdot j = - 2 z b \cdot A
\ee
and
\bea
\label{dlambda}
\partial_\mu \lambda &=& - {1 \over {\rm vol}(S^{d-1})} \int d^dx' \, \delta(\sigma z') \, {1 \over z} (x - x')_\mu \, 2 b \cdot j \\
&=& - 2x_\mu b \cdot A + {1 \over {\rm vol}(S^{d-1})} \int d^dx' \, \delta(\sigma z') \, {1 \over z} x'_\mu \, 2 b \cdot j
\eea
The gauge transformation restores holographic gauge, $A_z' = 0$, while combining (\ref{Amuprime}) and (\ref{dlambda}) we find
\bea
\big(zA_\mu\big)' &=& z A_\mu - 2 z b_\mu x \cdot A + {1 \over {\rm vol}(S^{d-1})} \int d^dx' \, \delta(\sigma z') \, x'_\mu \, 2 b \cdot j \\
\label{zAprime}
&=& z A_\mu + {1 \over {\rm vol}(S^{d-1})} \int d^dx' \, \delta(\sigma z') \, 2(x'_\mu \, b \cdot j - b_\mu x \cdot j)
\eea
Current conservation implies $\int d^dx' \, \theta(\sigma z') \, \partial_\mu j^\mu = 0$, which after integrating by parts means
\be
\int d^dx' \, \delta(\sigma z') \, (x - x')_\mu j^\mu = 0\,.
\ee
So we can replace $x$ with $x'$ in the last term of (\ref{zAprime}) to obtain
\be
\label{zAprime2}
\big(zA_\mu\big)' = z A_\mu + {1 \over {\rm vol}(S^{d-1})} \int d^dx' \, \delta(\sigma z') \, 2(x'_\mu \, b \cdot j - b_\mu x' \cdot j)
\ee
This establishes how the left hand side of (\ref{gaugesmear2}) behaves under a special conformal transformation.  Now let's look at the right hand side.
Under a special conformal transformation
\be
x^\mu \rightarrow x'{}^\mu = x^\mu + 2 b \cdot x x^\mu - b^\mu x^2
\ee
a vector of dimension $\Delta$ transforms according to
\be
\label{vector}
j_\mu \rightarrow j_\mu' = j_\mu + 2 x_\mu b \cdot j - 2 b_\mu x \cdot j - 2 \Delta b \cdot x j_\mu
\ee
The measure $d^dx' \, \delta(\sigma z')$ has dimension $1-d$ and transforms according to
\be
\label{measure}
d^dx' \, \delta(\sigma z') \rightarrow d^dx' \, \delta(\sigma z') \big[1 - 2(1-d) b \cdot x\big]
\ee
Combining (\ref{vector}) and (\ref{measure}) for $\Delta = d - 1$ reproduces the transformation seen in (\ref{zAprime2}).

This shows explicitly that the smearing function we have defined behaves covariantly under conformal transformations.  Indeed it seems that, aside from the
freedom to choose a different gauge in the bulk, the smearing function is uniquely fixed by the requirement of AdS covariance, at least if
one works on the complexified boundary.  This means that, even though we derived the smearing function by solving Maxwell's equations, it actually
has a more general scope of validity.  It can be used whenever one seeks a
linear map from a conserved current on the boundary to a gauge
field in the bulk.

\subsection{Two-point functions and bulk causality for gauge fields\label{sect:gauge2point}}

In this section we use the smearing functions we have constructed
to study bulk locality and causality for gauge fields.  Since we are working at leading
order in the $1/N$ expansion of the CFT, we are restricted to studying
bulk physics at the level of two-point functions.  We consider two basic cases:
in section \ref{sect:CS} we consider Chern-Simons theory in AdS${}_3$, and in section
\ref{sect:Maxwell} we consider Maxwell theory in AdS${}_4$ and higher.

\subsubsection{Chern-Simons fields in AdS${}_3$\label{sect:CS}}

AdS${}_3$ is something of a special case, since a conserved current
in the CFT is dual to a Chern-Simons gauge field in the bulk \cite{Jensen:2010em}.
Fortunately we can still use our smearing functions in this context, since
they're essentially fixed by AdS covariance.

From the smearing function (\ref{gaugesmear}) we have
\be
z A_\mu(t,x,z) = {1 \over 2 \pi} \int_0^{2\pi} z d\theta \, j_\mu(t + z \sin \theta, x + i z \cos \theta)
\ee
It's convenient to introduce light-front coordinates $x^\pm = t \pm x$ and
write the AdS${}_3$ metric as
\[
ds^2 = {R^2 \over z^2} \left(- d\xp d\xm + dz^2\right)
\]
For concreteness consider a CFT with a right-moving abelian current $j_- = j_-(\xm)$.
We assume the left-moving current vanishes, $j_+ = 0$.  Then the only non-trivial smearing integral is
\[
A_-(\xp,\xm,z) = \int_0^{2\pi} {d\theta \over 2 \pi} \, j_-(\xm - i z e^{i\theta})
\]
Defining $\xi = e^{i \theta}$ the contour integral picks up the pole at $\xi = 0$ and gives $A_-(\xp,\xm,z) = j_-(\xm)$.  So
a right-moving current in the CFT is dual to a bulk gauge field
\bea
\nonumber
&& A_+ = 0 \\
\label{CSmap}
&& A_-(\xp,\xm,z) = j_-(\xm) \\
\nonumber
&& A_z = 0
\eea
This is the world's simplest example of holography: the boundary current is lifted to be $z$-independent, and declared
to be a gauge field in the bulk.

Although ``reading the hologram'' in this case is almost trivial, there are a few things to check.  First of all, (\ref{CSmap}) defines
a flat gauge field in AdS, which satisfies the Chern-Simons equations of motion.\footnote{The smearing functions were constructed by solving Maxwell's equations, but they are essentially fixed by AdS covariance and therefore
hold more generally.  In AdS${}_3$ the smearing functions seem to know that a current in the CFT is
dual to a Chern-Simons gauge field in the bulk.}  Working backwards, the boundary conditions on the gauge field are
a bit different from (\ref{Fbc}), since we have
\[
A_\mu(x,z) \sim j_\mu(x) \qquad \hbox{\rm as $z \rightarrow 0$}
\]

We can use this framework to compute 2-point functions in the bulk.  The boundary correlator is fixed by conformal invariance.
With a Wightman $i \epsilon$ prescription
\be
\label{current-current}
\langle j_-(\xm) j_-(\xmp) \rangle = - {k \over 8 \pi^2} \, {1 \over (\xm - \xmp - i \epsilon)^2}
\ee
where $k$ is the level of the current algebra.  This lifts to a bulk correlator
\[
\langle A_-(\xp,\xm,z) A_-(\xpp,\xmp,z') \rangle = - {k \over 8 \pi^2} \, {1 \over (\xm - \xmp - i \epsilon)^2}
\]
Note that the bulk 2-point function is independent of $\xp$ and $z$, which is perhaps not so surprising in a topological theory.

We can also study bulk locality and causality in this framework.  The correlator (\ref{current-current}) implies that the
CFT currents obey the standard current algebra
\[
i [ j_-(\xm), j_-(\xmp) ] = - {k \over 4 \pi} \delta'(\xm - \xmp)\,.
\]
This lifts to a bulk commutator
\be
\label{CSbulk}
i [ A_-(\xp,\xm,z), A_-(\xpp,\xmp,z') ] = - {k \over 4 \pi} \delta'(\xm - \xmp)
\ee
This bulk commutator is clearly non-local, being independent of both $\xp$ and $z$.  But causality is respected: the field strength vanishes, so all
local gauge-invariant quantities obey causal (in fact trivial) commutation relations.

We obtained these results by applying our smearing functions to the current algebra on the boundary.  In appendix \ref{appendix:CS}
we show that they can also be obtained from the bulk point of view, by quantizing Chern-Simons theory in holographic gauge.

\subsubsection{Maxwell fields in AdS${}_4$ and higher\label{sect:Maxwell}}

We now consider Maxwell fields in AdS${}_4$ and higher, where a bulk gauge
field obeying Maxwell's equations is dual to a conserved current on the boundary.\footnote{Low
dimensions are special, for example in AdS${}_3$ a bulk Maxwell field is dual to a
gauge field in the CFT \cite{Marolf:2006nd,Jensen:2010em}.  Strictly speaking AdS${}_4$ Maxwell is also special since
the boundary currents only capture the ``electric''
sector of the bulk theory \cite{Witten:2003ya}.}

\begin{figure}
\begin{center}
\includegraphics[width=6.2cm]{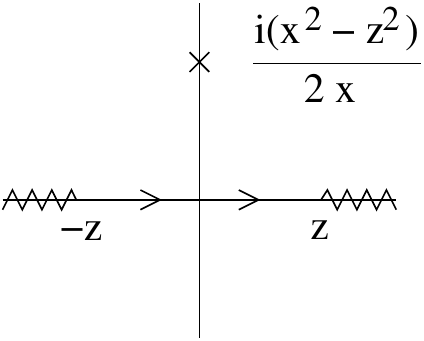}
\end{center}
\caption{Integration contour for $I_n$.  At large spacelike separation the pole is far up the imaginary axis.  The pole moves down and crosses the
integration contour when $x = z$; one can continue to smaller values of $x$ by deforming the contour.  The integral may be singular
when $x \rightarrow 0^+$ and the pole moves to $-i \infty$.  There are singularities when $x \rightarrow \pm i z$ and the pole hits an endpoint of the integration
contour.\label{fig:continue}}
\end{figure}

Our starting point is the current -- current correlator in a $d$-dimensional
CFT,
\be
\label{CFTcorrelator}
\langle \, j_\mu(x) \, j_\nu(0) \, \rangle = \left({1 \over x^2}\right)^{d-1} \left(\eta_{\mu\nu} - {2 x_\mu x_\nu \over x^2}\right)\,.
\ee
Up to an overall normalization, this correlator is fixed
by current conservation and conformal invariance.  We will be interested in Wightman correlators, defined by the $i \epsilon$ prescription
\[
x^2 \equiv - (t - i \epsilon)^2 + \vert {\bf x} \vert^2\,.
\]
Our goal is to apply the smearing function (\ref{gaugesmear}) to the first operator in (\ref{CFTcorrelator}),
to obtain a bulk - boundary correlator
\[
\langle \, A_\mu(t,{\bf x},z) \, j_\nu(0) \, \rangle\,.
\]
To deal with the vector indices it's useful to write the current -- current correlator in the form
\[
\langle \, j_\mu(x) \, j_\nu(0) \, \rangle = {d-2 \over d-1} \, \eta_{\mu\nu} \left({1 \over x^2}\right)^{d-1} - {1 \over 2(d-1)(d-2)} \,
\partial_\mu\partial_\nu \left({1 \over x^2}\right)^{d-2}
\]
Applying the smearing function (\ref{gaugesmear}) gives the bulk -- boundary correlator in terms of two scalar integrals,
\be
\label{MaxwellI1I2}
\langle z A_\mu(t,{\bf x},z) j_\nu(0) \rangle = {\Gamma(d/2) \over 2 \pi^{d/2}} \left(
{d-2 \over d-1} \, \eta_{\mu\nu} I_1 - {1 \over 2(d-1)(d-2)} \, \partial_\mu \partial_\nu I_2\right)
\ee
where
\be
\label{Indef}
I_n = \hspace{-7mm} \int\limits_{\hspace{8mm}t'{}^2 + \vert {\bf y}'\vert^2 = z^2}
\hspace{-5mm} dt' d^{d-1} y' \, {1 \over \big(-(t+t')^2 + \vert {\bf x} + i {\bf y}' \vert^2 \big)^{d-n}}
\ee
The integral is over a $(d-1)$-sphere of radius $z$ on the boundary.  We write the metric on
this sphere as
\[
ds^2 = {z^2 \over z^2 - y^2} \, dy^2 + (z^2 - y^2) \, d\Omega_{d-2}^2
\]
Here $-z < y < z$ and $d\Omega_{d-2}^2$ is the metric on a unit $S^{d-2}$.  To take advantage of spherical symmetry on $S^{d-2}$ we work at spacelike separation in the $x_1$ direction, setting
\[
x_1 = x \qquad t = x_2 = \cdots = x_{d-1} = 0
\]
Then $I_n$ reduces to a one-dimensional integral.
\[
I_n = {2 \pi^{(d-1)/2} \over \Gamma((d-1)/2)} \int_{-z}^z dy \,
{z (z^2 - y^2)^{(d-3)/2} \over (x^2 - z^2 + 2 i x y)^{d-n}}
\]
The prescription for defining this integral is to begin at large spacelike separation,
$x \gg 0$, where the operators are well-separated on the boundary and the integral is well-defined.
It can be extended to smaller values of $x$ by analytic continuation, as described in Fig.\ \ref{fig:continue}.  This prescription gives $I_n$ in terms
of a hypergeometric function.
\be
\label{In}
I_n = {2 \pi^{d/2} \over \Gamma(d/2)} \, {z^{d-1} \over (x^2)^{d-n}} F\Big(d-n,{d \over 2} - n + 1,
{d \over 2}, - {z^2 \over x^2} \Big)
\ee
When $n = 1$ this reduces to
\be
\label{I1}
I_1 = {2 \pi^{d/2} \over \Gamma(d/2)} \, {z^{d-1} \over (x^2 + z^2)^{d-1}}\,.
\ee
Note that $I_1$ is only singular on the bulk lightcone, at $x^2 + z^2 = 0$.  It
has an AdS-covariant form, with $I_1 \sim 1/(\sigma z')^{d-1}$.  These properties
could have been anticipated since, up to an overall coefficient, $I_1$ is the bulk - boundary correlator for a scalar field
with dimension $\Delta = d - 1$.

We are also interested in $n = 2$.  In any given dimension $I_2$ can be reduced to elementary functions, see for example Table \ref{table:I2}, however the expressions become unwieldy as $d$ increases.  For our purposes a key observation is that $I_2$ is singular on the boundary lightcone, with
\[
I_2 \sim {\pi^{(d+1)/2} \over 2^{d-4} \Gamma((d-1)/2)} \, {z \over x^{d-2}} \qquad
\hbox{\rm as $x \rightarrow 0$}
\]
$I_2$ is also singular on the bulk lightcone, at $x^2 + z^2 = 0$.

\begin{table}
\begin{center}
\begin{tabular}{l|l}
$d$ \quad & \qquad\quad $I_2$ \\
\hline
3 & \quad $\displaystyle - {2 \pi i z \over x} \log {x + i z \over x - i z}\phantom{\Bigg]}$ \\[10pt]
4 & \quad $\displaystyle {2 \pi^2 z^3 \over x^2 (x^2 + z^2)}$ \\[10pt]
5 & \quad $\displaystyle - {i \pi^2 z \over 2 x^3} \log {x + i z \over x - i z}
- {\pi^2 z^2 (x^2 - z^2) \over x^2 (x^2 + z^2)^2}$ \\[10pt]
6 & \quad $\displaystyle {\pi^3 z^5 (z^2 + 3 x^2) \over 3 x^4 (x^2 + z^2)^3}$
\end{tabular}
\end{center}
\caption{$I_2$ in various dimensions.\label{table:I2}}
\end{table}

Bulk -- boundary correlators follow from (\ref{MaxwellI1I2}) and (\ref{In}).  For example in AdS${}_4$ we find
\beas
\langle \, A_\mu(t,{\bf x},z) \, j_\nu(0) \, \rangle & = &
\eta_{\mu\nu}\left[{z(3x^2+z^2) \over 4x^2(x^2+z^2)^2} - {i \over 8 x^3} \log{x+ i z \over x - iz}\right] \\
& & \hspace{-6mm} - x_\mu x_\nu \left[{z(5x^2 + 3z^2) \over 4 x^4 (x^2 + z^2)^2} - {3 i \over 8 x^5} \log {x + i z \over x - i z}\right]
\eeas
while in AdS${}_5$ we have
\[
\langle \, A_\mu(t,{\bf x},z) \, j_\nu(0) \, \rangle =
\eta_{\mu\nu} \, {z^2(6 x^4 + 3 x^2 z^2 + z^4) \over 6 x^4 (x^2 + z^2)^3}
- x_\mu x_\nu \, {2 z^2 (3 x^4 + 3 x^2 z^2 + z^4) \over 3 x^6 (x^2 + z^2)^3}
\]
Explicit expressions in higher dimensions become rather unwieldy.  In general the $A$ -- $j$ correlators inherit the singularity structure of $I_2$:
they are singular on the boundary lightcone $x^2 = 0$, as well as on the bulk lightcone $x^2 + z^2 = 0$.  Correlators involving field strengths are both
simpler and better behaved.  In any dimension we find
\bea
\label{FieldStrength}
&& \langle \, F_{\lambda \mu}(t,{\bf x},z) \, j_\nu(0) \, \rangle = - {2(d-2)z^{d-2} \over (x^2 + z^2)^d} \, \left(x_\lambda \eta_{\mu\nu} - x_\mu \eta_{\lambda\nu}\right) \\
\nonumber
&& \langle \, F_{z\mu}(t,{\bf x},z) \, j_\nu(0) \, \rangle = {(d-2)z^{d-3} \over (x^2 + z^2)^d} \left(\eta_{\mu\nu} (x^2 - z^2) - 2 x_\mu x_\nu\right)
\eea
Note that $F$ -- $j$ correlators are only singular on the bulk lightcone.

Finally we can use these results to discuss bulk locality and causality.  The expectation value of a commutator $\langle \, [A_\mu(t,{\bf x},z),\, j_\nu(0)] \, \rangle$
is given by the difference in the prescriptions $t \rightarrow t - i \epsilon$ and $t \rightarrow t + i \epsilon$.  It follows that the commutator of a bulk gauge field
with a boundary current is non-zero at lightlike separation on the boundary.  Lightlike separation on the boundary implies spacelike
separation in the bulk, so we appear to have non-local or acausal correlators.  Of course there is no real violation of causality here, since $A$ -- $j$ correlators are gauge dependent.  For Maxwell fields we can
test causality by looking at gauge-invariant quantities, and indeed field strengths
have causal correlators: they commute with the boundary currents at bulk spacelike separation.

\section{Graviton smearing functions\label{sect:metric}}

We now turn our attention to constructing a smearing function that describes a fluctuation of the
bulk metric.  To this end we consider a linearized perturbation of the AdS metric,
\bea
\label{MetricPerturbation}
&& ds^2 = {R^2 \over z^2} \left(dz^2 + g_{\mu\nu} dx^\mu dx^\nu \right) \\
\nonumber
&& g_{\mu\nu} = \eta_{\mu\nu} + {z^2 \over R^2} h_{\mu\nu}
\eea
Here we are working in ``holographic gauge'' (or Fefferman - Graham coordinates \cite{FG}) in which
\[
g_{zz} = g_{z\mu} = 0
\]
The source-free Einstein equations in this coordinate system can be found in \cite{Henningson:1998gx}.\footnote{Ref.\ \cite{Henningson:1998gx}
uses $\rho = z^2/R^2$ as a radial coordinate.}  Working to linear order in $h_{\mu\nu}$ the $zz$, the $z\nu$, and the trace of the $\mu\nu$ components of
the Einstein equations read
\bea
\label{zz} && zz: \hspace{3mm} \Big(\partial_z^2 + {3 \over z} \partial_z \Big) h = 0 \\[2pt]
\label{znu} && z\nu: \hspace{3mm} \Big(\partial_z + {2 \over z} \Big)\left(\partial_\mu h^{\mu\nu} - \partial^\nu h \right) = 0 \\[2pt]
\label{trace} && {\rm trace}: \hspace{3mm} \Big(\partial_z^2 - {2d-5 \over z} \partial_z - {4 (d-1) \over z^2} \Big) h + 2 \big(\partial_\mu \partial^\mu h - \partial_\mu \partial_\nu h^{\mu\nu}\big) = 0 \qquad
\eea
Here $h \equiv h^\mu{}_\mu$.  The only solution to this system of equations compatible with normalizeable behavior as $z \rightarrow 0$ is to set\footnote{To see this
note that (\ref{znu}) implies $\partial_\mu h^{\mu\nu} - \partial^\nu h \sim 1/z^2$.  To avoid this non-normalizeable behavior we must set $\partial_\mu h^{\mu\nu} - \partial^\nu h = 0$.  The divergence of this equation means the last term in (\ref{trace}) drops out.  Then the difference of (\ref{zz}) and (\ref{trace}) gives $\Big(\partial_z + {2 \over z} \Big) h = 0$
which requires that we set $h = 0$.}
\be
\label{TracelessConserved}
h = 0 \qquad\quad \partial_\mu h^{\mu\nu} = 0
\ee
Thus $h_{\mu\nu}$ is traceless and conserved, which enables us to consistently identify its boundary behavior with the stress tensor of the CFT.

It only remains to solve the $\mu\nu$ components of the Einstein equations, which given (\ref{TracelessConserved}) can be simplified to
\[
\left(\partial_\alpha \partial^\alpha + \partial_z^2 + {5 - d \over z} \partial_z - {2 (d-2) \over z^2} \right) h_{\mu\nu} = 0
\]
Following the procedure that worked for Maxwell fields, we define $\phi_{\mu\nu} = z^2 h_{\mu\nu}$ and find that\footnote{This amounts to working in a vielbein
basis, $h_{ab} = e_a{}^\mu e_b{}^\nu h_{\mu\nu}$ where $e_a{}^\mu = {z \over R} \delta_a{}^\mu$.}
\[
\left(\partial_\alpha \partial^\alpha + z^{d-1} \partial_z {1 \over z^{d-1}} \partial_z \right) \phi_{\mu\nu} = 0
\]
That is, each component of $\phi_{\mu\nu}$ obeys the massless scalar wave equation.  A massless scalar is dual to an operator of
dimension $\Delta = d$ in the CFT, and has the asymptotic fall-off
\[
\phi_{\mu\nu}(x,z) \sim z^d T_{\mu\nu}(x) \qquad \hbox{\rm as $z \rightarrow 0$}
\]
We identify $T_{\mu\nu}$ with the stress tensor of the CFT.  To reconstruct the bulk metric perturbation from the stress tensor
we use the scalar smearing function (\ref{ScalarSmear}) given in appendix \ref{sect:scalar}.  Setting $\Delta = d$, this gives
\bea
\label{GravitonSmear}
&& z^2 h_{\mu\nu}(t,{\bf x},z) = {1 \over {\rm vol}(B^d)}
\hspace{-7mm} \int\limits_{\hspace{8mm}t'{}^2 + \vert {\bf y}'\vert^2 < z^2} \hspace{-5mm} dt' d^{d-1} y' \,
T_{\mu\nu}(t + t', {\bf x} + i {\bf y}') \\
\nonumber
&& \hbox{\rm volume of a unit $d$-ball} = {\rm vol}(B^d) = {2 \pi^{d/2} \over d \Gamma(d/2)}
\eea
Thus the bulk metric perturbation is obtained by smearing the stress tensor over a ball of radius $z$ on the
complexified boundary.

\subsection{AdS covariance\label{sect:metriccovariance}}

It's instructive to check that the smearing function (\ref{GravitonSmear}) respects AdS covariance.  We will be somewhat brief,
since the steps are very similar to those in section \ref{sect:gaugecovariance}.
Covariance under Poincar\'e transformations of $x^\mu$ is manifest.  A dilation corresponds to
the bulk isometry
\[
x^\mu \rightarrow x'{}^\mu = \lambda x^\mu \qquad\quad z \rightarrow z' = \lambda z\,.
\]
Holographic gauge is preserved since $h'_{zz} = h'_{z\mu} = 0$, while the combination $z^2 h_{\mu\nu}$ which appears on the left hand side of (\ref{GravitonSmear})
transforms like a scalar.  This matches the behavior of the right hand side: the stress tensor has dimension $d$, while the measure $d^dx'$ has dimension $-d$.

Special conformal transformations are a little more involved.  A special conformal transformation corresponds to an infinitesimal bulk isometry
\beas
&& x^\mu \rightarrow x'{}^\mu = x^\mu + 2 b \cdot x x^\mu - b^\mu (x^2 + z^2) \\
&& z \rightarrow z' = z + 2 b \cdot x z
\eeas
Under this isometry
\bea
\nonumber
&& h'_{zz} = 0 \\
\label{hprime}
&& h'_{z\mu} = 2 z b^\alpha h_{\alpha \mu} \\
\nonumber
&& h'_{\mu\nu} = h_{\mu\nu} + 2 b^\alpha (x_\mu h_{\alpha\nu} + x_\nu h_{\alpha \mu}) - 2 x^\alpha (b_\mu h_{\alpha\nu} + b_\nu h_{\alpha \mu}) - 4 b \cdot x h_{\mu\nu} \qquad
\eea
Holographic gauge isn't preserved, so to restore it we make a compensating diffeomorphism $x^\mu \rightarrow x^\mu + \epsilon^\mu(x,z)$, under which
\beas
&& \delta h_{\mu\nu} = {R^2 \over z^2} \big(\partial_\mu \epsilon_\nu + \partial_\nu \epsilon_\mu\big) \\
&& \delta h_{z\mu} = {R^2 \over z^2} \partial_z \epsilon_\mu \\
&& \delta h_{zz} = 0
\eeas
The appropriate diffeomorphism is
\be
\epsilon^\mu = - {1 \over R^2 {\rm vol}(B^d)} \int d^dx' \, \theta(\sigma z') \sigma z z' \, 2 b_\alpha T^{\alpha\mu}
\ee
for which
\bea
\nonumber
&& \delta h_{z\mu} = - 2 z b^\alpha h_{\alpha \mu} \\
\label{deltah}
&& \delta h_{\mu\nu} = - {1 \over z^2 {\rm vol}(B^d)} \int d^dx' \, \theta(\sigma z') \, 2 b^\alpha (x-x')_\mu T_{\alpha\nu} + (\mu \leftrightarrow \nu) \quad
\eea
This restores holographic gauge.  Combining (\ref{hprime}) and (\ref{deltah}) we find
\be
\label{deltaz2h}
\big(z^2 h_{\mu\nu}\big)' = z^2 h_{\mu\nu} + {1 \over {\rm vol}(B^d)} \int d^dx' \, \theta(\sigma z') \,
\left[2b^\alpha x'_\mu T_{\alpha\nu} - 2 x^\alpha b_\mu T_{\alpha\nu} + (\mu \leftrightarrow \nu)\right]
\ee
Current conservation in the form $\int d^dx' \, \theta(\sigma z') \sigma z z' \, \partial_\mu T^{\mu\nu} = 0$ implies
\[
\int d^dx' \, \theta(\sigma z') \, (x - x')^\mu T_{\mu\nu} = 0
\]
This means we can replace $x^\alpha$ with $x'{}^\alpha$ in (\ref{deltaz2h}), to obtain the transformation of the left hand side of (\ref{GravitonSmear}).
The result exactly matches the transformation of the right hand side, since under a special conformal transformation
\[
T_{\mu\nu} \rightarrow T'_{\mu\nu} = T_{\mu\nu} + 2 b^\alpha (x_\mu T_{\alpha \nu} + x_\nu T_{\alpha\mu}) - 2 x^\alpha (b_\mu T_{\alpha \nu} + b_\nu T_{\alpha \mu})
- 2 d b \cdot x T_{\mu\nu}
\]
The last term cancels the transformation of the measure $d^dx' \, \theta(\sigma z')$.

\subsection{Two-point functions and bulk causality for gravity\label{sect:gravity2point}}

We now use the smearing functions we have constructed to compute 2-point functions
for the graviton.  We consider gravity in AdS${}_3$ in section \ref{sect:AdS3}, and gravity
in AdS${}_4$ and higher in section \ref{sect:higherAdS}.

\subsubsection{Gravity in AdS${}_3$\label{sect:AdS3}}

AdS${}_3$ is special because there is no propagating graviton \cite{Deser:1983nh}.  Rather the
bulk curvature is completely determined by the vacuum Einstein equations
\be
\label{Einstein}
R_{MN} = {\Lambda \over d - 1} G_{MN}
\ee
where the cosmological constant $\Lambda = -d(d-1)/R^2$.  This uniquely fixes the geometry.  So in AdS${}_3$ we expect the smearing function
to generate a metric perturbation which corresponds to an infinitesimal (but non-normalizeable) diffeomorphism
of the background AdS metric.

We work in light-front coordinates $x^\pm = t \pm x$ and write the perturbed AdS metric as
\be
\label{AdS3Metric}
ds^2 = {R^2 \over z^2} \big(dz^2 - dx^+ dx^- \big) + h_{\mu\nu} dx^\mu dx^\nu
\ee
From the smearing function (\ref{GravitonSmear}) we have for instance
\be
z^2 h_{--} = {1 \over \pi} \hspace{-7mm} \int\limits_{\hspace{8mm}t'{}^2 + y'{}^2 < z^2} \hspace{-5mm} dt' dy' \,
T_{--}(t + t', x + iy')
\ee
Since $T_{--}$ only depends on $x^-$ this becomes ($t' = r \sin \theta$, $y' = r \cos \theta$)
\be
z^2 h_{--} = {1 \over \pi} \int_0^z rdr \int_0^{2\pi} d\theta \, T_{--} \big(x^- - i r e^{i\theta}\big)
\ee
Defining $\xi = e^{i \theta}$ the contour integral picks up the pole at $\xi = 0$ and ends up giving $h_{--} = T_{--}$.
So at the linearized level a stress tensor in the CFT corresponds to a bulk metric perturbation
\bea
\nonumber
&& h_{--} = T_{--}(x^-) \\
\label{AdS3MetricPerturbation}
&& h_{++} = T_{++}(x^+) \\
\nonumber
&& h_{+-} = 0
\eea
This provides a remarkably simple example of holography: the boundary stress tensor is lifted to be $z$-independent
and re-interpreted as a metric perturbation in the bulk.  Not surprisingly, this is very reminiscent of the Chern-Simons correspondence (\ref{CSmap}).

We can use this to compute the bulk 2-point function for the graviton.  For instance the CFT 2-point function
\be
\label{CFT2point}
\langle T_{--}(x^-) \, T_{--}(x'{}^-) \rangle = {c \over 8 \pi^2} {1 \over (x^- - x'{}^- - i \epsilon)^4}
\ee
lifts to a bulk correlator
\[
\langle h_{--}(x^+,x^-,z) \, h_{--}(x'{}^+,x'{}^-,z') \rangle = {c \over 8 \pi^2} {1 \over (x^- - x'{}^- - i \epsilon)^4}
\]
Here we have used a Wightman $i\epsilon$ prescription and $c$ is the central charge of the CFT.

To study bulk locality and causality in this framework, note that the CFT correlator (\ref{CFT2point}) corresponds to a Virasoro algebra
\[
i[T_{--}(x^-), T_{--}(x'{}^-) ] = {c \over 24 \pi} \delta^{\prime\prime\prime}(x^- - x'{}^-)
\]
This lifts to the bulk commutator
\[
i[h_{--}(x^+,x^-,z), h_{--}(x'{}^+,x'{}^-,z') ] = {c \over 24 \pi} \delta^{\prime\prime\prime}(x^- - x'{}^-)
\]
Metric perturbations in the bulk have non-local commutators; this behavior is acceptable since metric perturbations are coordinate dependent.
One might ask if there is a quantity -- analogous to the field strength for a gauge field -- which obeys causal commutation relations.  In the next
section we will claim that, for gravity, such a quantity is provided by the Weyl tensor.  This claim becomes vacuous in three dimensions since
the Weyl tensor vanishes identically.

We began this section by recalling that the source-free Einstein equations fix the bulk geometry to be pure AdS.  So to complete the story,
one might ask for a coordinate transformation which brings the perturbed metric (\ref{AdS3Metric}), (\ref{AdS3MetricPerturbation}) back
to the canonical form $ds^2 = {R^2 \over z^2} \big(dz^2 - dx^+ dx^-\big)$.  The required transformation is
\bea
\nonumber
&& \delta x^+ = - {2 \over R^2} \, {1 \over \partial_+^3} T_{++} - {z^2 \over R^2} \, {1 \over \partial_-} T_{--} \\
&& \delta x^- = - {2 \over R^2} \, {1 \over \partial_-^3} T_{--} - {z^2 \over R^2} \, {1 \over \partial_+} T_{++} \\
\nonumber
&& \delta z = - {z \over R^2} \, \left({1 \over \partial_+^2} T_{++} + {1 \over \partial_-^2} T_{--}\right)
\eea
Note that the transformation does not vanish at the boundary, so it does not correspond to a (normalizeable) gauge symmetry of the bulk theory.

\subsubsection{Gravity in AdS${}_4$ and higher\label{sect:higherAdS}}

Our starting point for gravity in AdS${}_4$ and higher is the 2-point function of the stress tensor in a general CFT.  Up to an overall coefficient
proportional to the central charge, this has the form\footnote{See for example (2.37) and (A5) in Ref.\ \cite{Erdmenger:1996yc}.}
\be
\langle T_{\mu\nu}(x) T_{\alpha\beta}(0) \rangle = X_{\mu\nu\alpha\beta} {1 \over (x^2)^d} + Y_{\mu\nu\alpha\beta} {1 \over (x^2)^{d-1}}
+ Z_{\mu\nu\alpha\beta} {1 \over (x^2)^{d-2}}
\ee
where we've introduced
\bea
\label{XYZ}
&& X_{\mu\nu\alpha\beta} = -2d \, \eta_{\mu\nu} \eta_{\alpha\beta} + d(d-1)\big(\eta_{\mu\alpha} \eta_{\nu\beta} + \eta_{\mu\beta} \eta_{\nu\alpha} \big) \\
\nonumber
&& Y_{\mu\nu\alpha\beta} = {1 \over d-1} \big(\eta_{\mu\nu} \partial_\alpha \partial_\beta + \eta_{\alpha\beta} \partial_\mu \partial_\nu\big)
- {1 \over 2} \big(\eta_{\mu\alpha} \partial_\nu \partial_\beta + \eta_{\mu\beta} \partial_\nu \partial_\alpha + \eta_{\nu\alpha} \partial_\mu \partial_\beta
+ \eta_{\nu\beta} \partial_\mu \partial_\alpha \big) \\
\nonumber
&& Z_{\mu\nu\alpha\beta} = {1 \over 2 (d-1)(d-2)} \partial_\mu \partial_\nu \partial_\alpha \partial_\beta
\eea
Up to an overall normalization this correlator is uniquely determined by requiring that the stress tensor be traceless
and conserved with the correct scaling dimension.  Applying the smearing function (\ref{GravitonSmear}) gives the bulk -- boundary correlator
\be
\label{hT}
z^2 \langle h_{\mu\nu}(t,{\bf x},z) T_{\alpha\beta}(0) \rangle = X_{\mu\nu\alpha\beta} J_0 + Y_{\mu\nu\alpha\beta} J_1 + Z_{\mu\nu\alpha\beta} J_2
\ee
where
\be
J_n = {1 \over {\rm vol}(B^d)} \hspace{-7mm} \int\limits_{\hspace{8mm}t'{}^2 + \vert {\bf y}'\vert^2 < z^2} \hspace{-5mm} dt' d^{d-1} y' \,
{1 \over \big(-(t+t')^2 + \vert {\bf x} + i {\bf y}' \vert^2 \big)^{d-n}}
\ee
Note that $J_n$ is related to the integral (\ref{Indef}) we encountered for gauge fields,
\[
{d \over dz} J_n = {1 \over {\rm vol}(B^d)} I_n\,.
\]
Integrating (\ref{In}) gives
\be
\label{Jn}
J_n = {z^d \over (x^2)^{d-n}} F\Big(d-n,{d \over 2} - n + 1, {d \over 2} + 1, - {z^2 \over x^2}\Big)
\ee
In general $J_n$ has singularities on both the boundary lightcone (where $x^2 = 0$) and the bulk lightcone (where $x^2 + z^2 = 0$).
The case $n = 0$ is an exception to this general rule, since
\[
J_0 = {z^d \over (x^2 + z^2)^d}
\]
$J_0$ is only singular on the bulk lightcone, and in fact has an AdS-covariant form $J_0 \sim 1 / (\sigma z')^d$.  This was to be expected since,
up to an overall normalization, $J_0$ is the bulk -- boundary correlator for a massless scalar field.  Some other cases of interest can be found
in table \ref{table:J1J2}.

\begin{table}
\begin{center}
\hspace*{-8mm}\begin{tabular}{l|c|c}
$d$ \quad & $J_1$ & $J_2$ \\
\hline
3 & $\displaystyle - {3 i \over 4 x} \log {x + i z \over x - i z} - {3 z \over 2(x^2 + z^2)} \phantom{\Bigg]}$
   & $\displaystyle - {3 i (x^2 + z^2) \over 4 x} \log {x+iz \over x-iz} - {3 z \over 2}$ \\[10pt]
4 & $\displaystyle {z^4 \over x^2 (x^2 + z^2)^2}$
   & $\displaystyle -2 \log {x^2 + z^2 \over x^2} + {2 z^2 \over x^2}$ \\[10pt]
5 & $\displaystyle - {5i \over 32 x^3} \log {x + i z \over x - i z} - {5 z (3 x^4 + 8 x^2 z^2 - 3 z^4) \over 48 x^2 (x^2 + z^2)^3}$
   & $\displaystyle {15 i \over 32 x^3} (3x^2 - z^2) \log {x+iz \over x - iz} + {15 z (3x^2 + z^2) \over 16 x^2 (x^2 + z^2)}$ \\[10pt]
6 & $\displaystyle {z^6 (4 x^2 + z^2) \over 4 x^4 (x^2 + z^2)^4}$
   & $\displaystyle {z^6 \over x^4 (x^2 + z^2)^2}$
\end{tabular}
\end{center}
\caption{$J_1$ and $J_2$ in low dimensions.\label{table:J1J2}}
\end{table}

At this stage we have an expression for the $h$ -- $T$ correlator in terms of differential operators acting on $J_n$'s.  We will stop here, since explicitly
evaluating the derivatives in (\ref{hT}) leads to lengthy expressions.  But one important observation we can make is that the $h$ -- $T$ correlator inherits the
singularity structure of $J_1$ and $J_2$: it has singularities on both the bulk and boundary lightcones.  This means the commutator $[ h_{\mu\nu}(t,{\bf x},z),
T_{\alpha\beta}(0) ]$ will be non-zero at lightlike separation on the boundary (where $x^2 = 0$), even though this corresponds to spacelike separation in the bulk
(since $x^2 + z^2 > 0$).  This shows that in holographic gauge metric perturbations have acausal commutators.  This is acceptable because the commutator is gauge dependent.

This raises an interesting question, whether there is a quantity one can define in linearized gravity which obeys causal commutation relations.  That is, whether there is
something analogous to the Maxwell field strength $F_{\mu\nu}$, which as we saw in (\ref{FieldStrength}) has correlators that are only singular on the bulk lightcone.
At first one might think the gravitational analog is provided by the Riemann tensor.  However this can't be right: perturbing the source-free Einstein equations (\ref{Einstein})
shows that $\delta R_{\mu\nu} = - {d \over R^2} h_{\mu\nu}$.  Since we've already shown that the metric perturbation has acausal commutators, the same must be
true for the Ricci tensor.

This suggests that we split off the Ricci part of the curvature and work with the Weyl tensor.  In fact the Weyl tensor commutes with the
boundary stress tensor at bulk spacelike separation.  We will show this in two ways: first by an intuitive argument, then by an
explicit calculation in holographic gauge.

The intuitive argument runs as follows.  Imagine quantizing the bulk theory perturbatively using a covariant gauge condition.  Then locality would be manifest, and
all fields (including the metric perturbation) would obey canonical local commutation relations.  It follows that in covariant gauge the Weyl tensor commutes with the boundary
stress tensor at spacelike separation.  But since the Weyl tensor transforms homogeneously under changes of coordinates, if the commutator vanishes in covariant gauge
it should also vanish in holographic gauge.\footnote{This argument breaks down for the Riemann tensor.  In an AdS background the Riemann tensor acquires a vev,
and a perturbation $\delta R_{\alpha\beta\gamma\delta}$ transforms inhomogeneously under changes of coordinates.  By contrast the Weyl tensor has a vanishing vev and
transforms homogeneously.  It follows that, at the linearized level,
the Weyl tensor is gauge invariant around an AdS background.}

The explicit calculation proceeds as follows.  Linearizing around an AdS background the non-trivial components of the Weyl tensor are
\bea
\label{Weyl1}
z^2 C_{\alpha\beta\gamma\delta} & = & {1 \over 2} \big(\partial_\alpha \partial_\gamma \phi_{\beta\delta} - \partial_\alpha \partial_\delta \phi_{\beta\gamma}
- \partial_\beta \partial_\gamma \phi_{\alpha\delta} + \partial_\beta \partial_\delta \phi_{\alpha\gamma}\big) \\
\nonumber
&&- {1 \over 2z} \partial_z \big(\eta_{\alpha\gamma} \phi_{\beta\delta} - \eta_{\alpha\delta} \phi_{\beta\gamma} - \eta_{\beta\gamma} \phi_{\alpha\delta}
+ \eta_{\beta\delta} \phi_{\alpha\gamma}\big) \\[5pt]
\nonumber
z^2 C_{z\beta\gamma\delta} & = & {1 \over 2} \partial_z \big(\partial_\gamma \phi_{\beta\delta} - \partial_\delta \phi_{\beta\gamma}\big)
\eea
Here $\phi_{\alpha\beta} = z^2 h_{\alpha\beta}$, and we have used the fact that $\phi_{\alpha\beta}$ obeys the massless scalar wave equation
$\big(\partial_\alpha \partial^\alpha + \partial_z^2\big)\phi_{\mu\nu} = {d - 1 \over z} \partial_z \phi_{\mu\nu}$.  The remaining components of the Weyl
tensor $C_{z\beta z\delta}$ are not independent by the trace-free condition.

In principle it is straightforward to compute $C$ -- $T$ correlators.  Consider for example $z^2 \langle C_{z\beta\gamma\delta}(x) T_{\rho\sigma}(0) \rangle$.
Using the $\phi$ -- $T$ correlator (\ref{hT}) and the operators (\ref{XYZ}) one obtains a rather long expression.  However many terms drop out
when antisymmetrized on $\gamma$ and $\delta$.  What survives has the form (`stuff' meaning metrics and derivatives tangent to the boundary)
\bea
\nonumber
z^2 \langle C_{z\beta\gamma\delta}(x) T_{\rho\sigma}(0) \rangle &=& \partial_z \int_{x'{}^2 < z^2} d^dx' \,
\left\lbrace ({\rm stuff}) \cdot {1 \over (x^2)^d} + ({\rm stuff}) \cdot {1 \over (x^2)^{d-1}} \right\rbrace \\
\nonumber
&=& \int_{x'{}^2 = z^2} d^dx' \, \left\lbrace ({\rm stuff}) \cdot {1 \over (x^2)^d} + ({\rm stuff}) \cdot {1 \over (x^2)^{d-1}} \right\rbrace \\[5pt]
\label{CzT}
&=& ({\rm stuff}) \cdot I_0 + ({\rm stuff}) \cdot I_1
\eea
As we saw in (\ref{I1}) $I_1$ is analytic on the boundary lightcone.  It turns out that $I_0$ is also analytic at $x^2 = 0$:
\be
\label{I0}
I_0 = {2 \pi^{d/2} \over \Gamma(d/2)} \, {z^{d-1} (x^2 - z^2) \over (x^2 + z^2)^{d+1}}
\ee
So the correlator (\ref{CzT}) is analytic at $x^2 = 0$, and $C_{z\beta\gamma\delta}$ obeys causal commutation
relations with the boundary stress tensor.

Now consider $z^2 \langle C_{\alpha\beta\gamma\delta}(x) T_{\rho\sigma}(0) \rangle$.  Again one obtains
a rather long expression.  However many terms drop out when antisymmetrized on $\alpha$ and $\beta$, or on $\gamma$ and $\delta$.  Also many
terms involve either $J_0$, $I_0$ or $I_1$ which we know are analytic at $x^2 = 0$.  Dropping all such contributions, up to an overall coefficient we find that
only two terms survive:
\bea
&& z^2 \langle C_{\alpha\beta\gamma\delta}(x) T_{\rho\sigma}(0) \rangle \\
\nonumber
&\sim& \partial_{[\alpha} \eta_{\beta][\gamma} \partial_{\delta]} \partial_\rho \partial_\sigma
\Big(\int_{x'{}^2 < z^2} d^dx' \, {1 \over (x^2)^{d-1}} - {1 \over 2(d-2)z} \int_{x'{}^2 = z^2} d^dx' \, {1 \over (x^2)^{d-2}}\Big)
\eea
With the help of one of Gauss' recursion relations for hypergeometric functions one can show that the quantity in parenthesis is
\[
{\rm vol}(B^d) J_1 - {1 \over 2 (d-2) z} I_2 = - {\pi^{d/2} \over (d-2) \Gamma(d/2)} \, {z^{d-2} \over (x^2 + z^2)^{d-2}}
\]
This is analytic on the boundary lightcone, so $C_{\alpha\beta\gamma\delta}$
obeys causal commutation relations with the boundary stress tensor.

\section{Massive vector fields\label{sect:massivevector}}

In this section we derive the smearing function for a massive vector.  Our starting point is
the Lagrangian for a massive vector field in Lorentzian AdS${}_{d+1}$.
\begin{equation}
S=\int dz d^{d}x \sqrt{-G}\, \left(-\frac{1}{4}F^{M N}F_{M N}-\frac{1}{2}m^2 A_{M}A^{M}\right)
\end{equation}
The equations of motion $\nabla_{M}F^{M N}-m^2A^{N}=0$ imply
\begin{equation}
\nabla_{M} A^{M}=0\,.
\label{relation}
\end{equation}
Decomposing $A_M = (A_z,A_\mu)$, the equations of motion for $A_z$ are
\begin{equation}
\left(\partial^{2}_{z}+\partial_{\mu}\partial^{\mu}-\frac{1}{z}(d-1)\partial_{z}-\frac{m^2-d+1}{z^2}\right)A_{z}=0
\label{az}
\end{equation}
This is identical to the equation of motion for a scalar field with $({\rm mass})^2 = m^2-d+1$.
For the other components one has (defining $\phi_{\mu}=zA_{\mu}$)
\begin{equation}
\left(\partial^{2}_{z}+\partial_{\nu}\partial^{\nu}-\frac{1}{z}(d-1)\partial_{z}-\frac{m^2-d+1}{z^2}\right)\phi_{\mu}=2\partial_{\mu} A_{z}
\label{ai}
\end{equation}
Let
\begin{equation}
 \Delta=\frac{d}{2}+\sqrt{\frac{(d-2)^2}{4}+m^2}
 \end{equation}
 and define the boundary value of $A_{z}$ by
 \[
 A_{z} \sim z^{\Delta}A_{z}^{0} \qquad \hbox{\rm as $z \rightarrow 0$}
 \]
 The equation of motion for $A_{z}$ can be solved in the same way as for a scalar field (see appendix \ref{sect:scalar})
 \begin{equation}
 A_{z}(t,{\bf x},z)= \int_{t^{'2} + {\bf y}^{'2} < z^2 }dt' d{\bf y}' \, \left(\frac{z^2-t^{'2}-{\bf y}^{'2}}{z}\right)^{\Delta-d}A_{z}^{0}(t+t',{\bf x}+i{\bf y}')
 \end{equation}
 What is the boundary value of $A_{z}^{0}$ in terms of CFT data?  Since $\phi_{\mu}(z\rightarrow 0)\sim z^{\Delta}$ then $A_{\mu} \sim z^{\Delta-1}j_{\mu}$, and
 inserting this in (\ref{relation}) gives
 \begin{equation}
 A_{z}^{0}=\frac{1}{d-\Delta -1} \, \partial_{\mu} j^{\mu}
 \end{equation}
 So $A_z^0$ is sourced by the divergence of the boundary current.
 
 Now let's solve (\ref{ai}). First note that a solution to the homogeneous equation (\ref{az}) can be expanded in modes as
 \begin{equation}
 A_z=\int_{|\omega|>|k|} d\omega d^{d-1}k \, a^{z}_{\omega k}e^{-i\omega t+i{\bf k}{\bf x}}z^{d/2}J_{\nu}(z\sqrt{\omega^2-{\bf k}^2})
 \end{equation}
 where $\nu=\Delta-d/2$ and $J_{\nu}(y)$ is a Bessel function.
 A similar solution would hold for (\ref{ai}) if the right hand side was zero.
 The complete solution to (\ref{ai}) can then be written in the form \cite{Mueck:1998iz}
 \begin{eqnarray}
  \label{solution}
  \phi_\mu(t,{\bf x},z) &= &\int_{|\omega|>|k|} d\omega d^{d-1}k \, z^{d/2}e^{-i\omega t+i{\bf k}{\bf x}} \\
  \nonumber
  &\times &\left(a^{\mu}_{\omega k}J_{\nu}(z\sqrt{\omega^2-{\bf k}^2})+a^{z}_{\omega k}\frac{izk_{\mu}}{\sqrt{\omega^2-{\bf k}^2}}J_{\nu+1}(z\sqrt{\omega^2-{\bf k}^2})\right)
 \end{eqnarray}
 Now from the boundary behavior of $A_{z}$ one has
 \begin{equation}
 a^{z}_{\omega k}=\frac{2^{\nu}\Gamma (\nu+1)}{(2\pi)^d (\omega^2-{\bf k}^2)^{\nu/2}}\int dt' d^{d-1}x'
 e^{i \omega t'-i{\bf k}{\bf x}'}A_{z}^{0}(t',{\bf x}')
 \end{equation}
 and since the term proportional to $a^{z}_{\omega k}$ in (\ref{solution}) is subleading as $z \rightarrow 0$ one also has
 \begin{equation}
 a^{\mu}_{\omega k}=\frac{2^{\nu}\Gamma (\nu+1)}{(2\pi)^d (\omega^2-{\bf k}^2)^{\nu/2}}\int dt' d^{d-1}x'
 e^{i \omega t'-i{\bf k}{\bf x}'}z j_{\mu}(t',{\bf x}')
 \end{equation}
 By inserting the expressions for $a^{\mu}_{\omega k}$ and $a^{z}_{\omega k}$ into (\ref{solution}) one gets an expression for the bulk field in terms of boundary data.
 The first term looks just like the smearing function for a scalar field of dimension $\Delta$, while the second term (aside from a factor $\frac{izk_\mu}{2(\nu+1)}$) is just the smearing function for a scalar field of dimension $\Delta+1$ \cite{Hamilton:2006fh}.  As a result we get the following expression
 \begin{equation}
 \label{MassiveVectorSmear}
 \phi_{\mu}(t,{\bf x},z)=\int K_{\Delta}(x,x')j_{\mu}(x') +\frac{z}{2(\nu+1)}\int K_{\Delta+1}(x,x') \, \partial_{\mu}A_{z}^{0}(x')
 \end{equation}
 More explicitly
 \begin{eqnarray}
 &&zA_{\mu}(t,{\bf x},z)=\frac{\Gamma(\Delta -d/2+1)}{\pi^{d/2}\Gamma(\Delta-d+1)}\int_{t'^2+{\bf y}'^2<z^2}dt'd^{d-1}y'\Big(\frac{z^2-t'^2 - {\bf y}'^2}{z}\Big)^{\Delta-d} A^{0}_{\mu}(t+t',x+i{\bf y}')\nonumber\\
 && \quad + \frac{z\Gamma(\Delta -d/2+1)}{2\pi^{d/2}\Gamma(\Delta -d+2)} \int_{t'^2+{\bf y}'^2<z^2}dt'd^{d-1}y'\Big(\frac{z^2-t'^2-{\bf y}'^2}{z}\Big)^{\Delta-d+1} \partial_\mu A_{z}^{0}(t+t',x+i{\bf y}') \nonumber
\end{eqnarray}

\subsection{Two-point functions and bulk causality}

In this section we compute the two point function of a massive vector.
The CFT two point function for a spin-1 field is
\begin{equation}
<j_{\mu}(x) j_{\nu}(0)>=\left(\eta_{\mu \nu}-\frac{2x_{\mu}x_{\nu}}{x^2}\right)\frac{1}{(x^2)^{\Delta}}
\label{amuanu}
\end{equation}
It can also be written in the form
\begin{equation}
<j_{\mu}(x) j_{\nu}(0)>=\frac{\Delta -1}{\Delta}\eta_{\mu \nu}\frac{1}{(x^2)^{\Delta}}-\frac{1}{2\Delta (\Delta-1)} \partial_{\mu}\partial_{\nu}\frac{1}{(x^2)^{\Delta-1}}
\label{amuanunew}
\end{equation}
Since our expression for the bulk operator involves the divergence of the current we will also need
\begin{equation}
<\partial_{\mu}j^{\mu} (x) j_{\nu}(0)>=\frac{d-\Delta -1}{\Delta} \partial_{\nu}\frac{1}{(x^2)^{\Delta}}
\end{equation}
The correlator of a bulk field $A_{z}$ with a boundary current $j_\nu$ is easy to read off from the smearing function for $A_{z}$,
which as we showed is just the smearing function of a scalar field of dimension $\Delta$. Since $A_{z}(x)=\frac{1}{d-\Delta-1}\partial_{\mu}j^{\mu} (x)$
we have
\begin{equation}
<A_{z}(z,x) j_{\nu}(0)>=\frac{1}{\Delta}\partial_{\nu}\left(\frac{z}{x^2+z^2}\right)^{\Delta}
\label{azanu}
\end{equation}
This two-point function respects bulk causality.

For the other components of the bulk field we have
\begin{eqnarray}
\label{twopoint}
&&<zA_{\mu}(t,{\bf x},z)j_{\nu}(0)> = \int_{t'^2+|{\bf y}'|^2<z^2}dt'd^{d-1}y' \bigg[ \\
\nonumber
&& \frac{\Gamma(\Delta -d/2+1)}{\pi^{d/2}\Gamma(\Delta-d+1)} \left(\frac{z^2-t'^2-|{\bf y}'|^2}{z}\right)^{\Delta-d}<j_{\mu}(t+t',{\bf x}+i{\bf y}') j_{\nu}(0)> \\
\nonumber
&&+\frac{z\Gamma(\Delta-d/2+1)}{2\pi^{d/2}\Gamma(\Delta-d+2)} \left(\frac{z^2-t'^2-|{\bf y}'|^2}{z}\right)^{\Delta-d+1}\partial_{\mu}<A_{z}(t+t',{\bf x}+i{\bf y}') j_{\nu}(0)>\bigg]
\end{eqnarray}
Using (\ref{amuanunew}) and (\ref{azanu}) we write this as
\begin{eqnarray}
& &<zA_{\mu}(x,z)j_{\nu}(0)>=\frac{\Delta-1}{\Delta}\eta_{\mu \nu}\left(\frac{z}{z^2+x^2}\right)^{\Delta} \nonumber \\
&-&\frac{\Gamma(\Delta-d/2+1)}{2 \Delta \pi^{d/2}\Gamma(\Delta-d+1)}\partial_{\mu}\partial_{\nu}\left(\frac{1}{\Delta-1}f_{\Delta}(z,x)-\frac{z}{\Delta-d+1}f_{\Delta+1}(z,x)\right)\nonumber
\end{eqnarray}
where
\begin{eqnarray}
f_{\Delta}(z,x)=\int_{t'^2+|y'|^2<z^2}dt'd^{d-1}y' \left(\frac{z^2-t'^2-|y'|^2}{z}\right)^{\Delta-d}\times\nonumber\\
\left(\frac{1}{-(t+t')^2+(x_{1}+iy_{1})^2+\cdots +(x_{d-1}+iy_{d-1})^2}\right)^{\Delta-1}\nonumber
\end{eqnarray}
 We set $t=0$, $x_1=x$, $x_2= \cdots = x_{d-1}=0$.
We will compute $f_{\Delta}$ for this case then restore the dependence on the other coordinates using Lorentz invariance.
Switching from $(t',y')$ to spherical coordinates we get
\begin{equation}
f_{\Delta}={\rm vol}(S^{d-2})\int_{0}^{z}dr r^{d-1}\left(\frac{z^2-r^2}{z}\right)^{\Delta-d}\int_{0}^{\pi}\frac{\sin^{d-2} \theta}{(x^2+2ixr\cos \theta-r^2)^{\Delta-1}}
\end{equation}
We use the integrals
\begin{eqnarray}
&&\int_{0}^{\pi}\frac{\sin^{2\mu-1} \theta}{(1+2a\cos \theta+a^2)^{\nu}}=\frac{\Gamma(\mu)\Gamma(\frac{1}{2})}{\Gamma(\mu+\frac{1}{2})}F(\nu,\nu-\mu+\frac{1}{2},\mu+\frac{1}{2},a^2)\nonumber \\
&&\int_{0}^{1}(1-x)^{\mu-1}x^{\gamma-1}F(\alpha,\beta, \gamma,ax)=\frac{\Gamma(\mu)\Gamma(\gamma)}{\Gamma(\mu+\gamma)}F(\alpha,\beta,\gamma+\mu,a)
\label{iden1}
\end{eqnarray}
to find
\begin{equation}
f_{\Delta}=\frac{\pi^{d/2}\Gamma(\Delta-d+1)}{\Gamma(\Delta-\frac{d}{2}+1)}\frac{z^{\Delta}}{x^{2\Delta -2}}F\Big(\Delta-1,\Delta-\frac{d}{2},\Delta-\frac{d}{2}+1,-\frac{z^2}{x^2}\Big)
\end{equation}
Then we use the identity
\begin{equation}
\gamma F(\alpha,\beta,\gamma,x)-\gamma F(\alpha,\beta+1,\gamma,x)+x\alpha F(\alpha+1,\beta+1,\gamma+1,x) = 0
\label{iden2}
\end{equation}
and restore Lorentz invariance to find
\begin{equation}
<zA_{\mu}(x,z)j_{\nu}(0)>=\frac{\Delta-1}{\Delta}\eta_{\mu \nu}\left(\frac{z}{x^2+z^2}\right)^{\Delta}-\frac{z^{\Delta}}{2\Delta(\Delta-1)}\partial_{\mu}\partial_{\nu}\left(\frac{1}{x^2+z^2}\right)^{\Delta-1}
\end{equation}
Note that the final answer is only non-analytic on the bulk lightcone. This however was achieved by a cancellation of terms that are non-analytic on the boundary lightcone between 
$f_\Delta$ and $f_{\Delta+1}$. So the locality of a massive vector field in the bulk is made possible by the fact that the dual boundary current isn't conserved, which allowed us
to cancel non-analytic terms in the correlator. This mechanism is not available for a gauge field since it is dual to a conserved current.

\section{Interactions\label{sect:interaction}}

In this section we make some remarks on constructing bulk operators at higher orders in $1/N$.  For scalar fields it was shown in
\cite{Kabat:2011rz} that one can construct interacting local bulk fields without any knowledge of the bulk Lagrangian.  Rather,
by adopting bulk micro-causality as a guiding principle, one can construct the appropriate bulk operators just from knowing CFT
correlators.  Here we show that something similar can be done for a massive vector
field in AdS${}_3$: a local bulk operator can be constructed, even in the presence of interactions.  However for a gauge field in AdS${}_3$ we show that the analogous procedure breaks down.  In this section, to avoid notational complexity, we denote
\beas
&& w = x^+ = t + x \\
&& \bar{w} = x^- = t - x
\eeas

Up to an overall coefficient, the three point function of three primary operators in a two dimensional CFT is
\bea
&& <{\cal O}_{1, h_1,\bar{h}_1}(w_1,\bar{w}_1){\cal O}_{2,h_2,\bar{h}_2}(w_2,\bar{w}_2){\cal O}_{3,h_3,\bar{h}_3}(w_3,\bar{w}_3)> \\
\nonumber
&=&
\frac{1}{w_{12}^{h_1 +h_2 -h_3}w_{23}^{h_2 +h_3 -h_1}w_{13}^{h_3+h_1-h_2}}
\frac{1}{\bar{w}_{12}^{\bar{h}_1 +\bar{h}_2 -\bar{h}_3}\bar{w}_{23}^{\bar{h}_2 +\bar{h}_3 -\bar{h}_1}\bar{w}_{13}^{\bar{h}_3+\bar{h}_1-\bar{h}_2}}
\eea
Here $w_{ij}=w_i-w_j$.
Let us for simplicity assume that ${\cal O}_{2}$ and ${\cal O}_{3}$ are scalar operators so $h_2=\bar{h}_2$ and $h_{3}=\bar{h}_{3}$, but ${\cal O}_{1}$ has spin 1 with $h_{1}=\bar{h}_1+1$. To explore bulk locality we smear ${\cal O}_{2}$ into a bulk operator using the free field smearing function
\begin{equation}
{\cal O}_{2}(z,w_2,\bar{w}_2)=\int_{0}^{z}rdr (\frac{z^2-r^2}{z})^{2h-2}\int_{|\alpha|=1}\frac{d\alpha}{i\alpha} {\cal O}(w_2 +r\alpha, \bar{w}_2 -r\alpha^{-1}).
\end{equation}
We can get the CFT three point function with $h_1\rightarrow h_1 +1$ (as long as $h_{1} \neq 0$) by acting on a three point correlator with the operator
\begin{equation}
\frac{1}{h_3-h_2-h_1}\frac{\partial}{\partial w_{12}}-\frac{1}{h_2-h_3-h_1}\frac{\partial}{\partial w_{13}}
\label{oper}
\end{equation}
So the result for $h_{1}=\bar{h}_1+1$ can be gotten from the result for $h_{1}=\bar{h}_{1}$ by acting with the operator (\ref{oper}). The situation with
 $h_{1}=\bar{h}_{1}$ was analyzed in \cite{Kabat:2011rz}. It was found that for scalar operators one can add
 a series of appropriately smeared higher dimension scalar operators that will cancel the causality-violating terms in the three point function. Here we see that this is still true if one of the boundary operator has spin. Note however that for the special case of conserved current (meaning $h=0$, $\bar{h}=1$ or $h=1$, $\bar{h}=0$) this argument does not apply. This is not only because acting with the operator (\ref{oper}) is not possible, but also because if ${\cal O}_{1}$ is a conserved current then Ward identities restrict its three point function. For instance for a conserved current the three point function will vanish unless the two point function $<{\cal O}_{2}{\cal O}_{3}>$ is non-zero. So for a conserved current adding smeared higher dimension primaries is not in general possible.
 
We now consider the case where ${\cal O}_{1}$ is smeared into the bulk. We'll work in terms of the OPE, similarly to what was done in \cite{Kabat:2011rz}.
For simplicity we denote $h_1=n$, $\bar{h}_1=n-1$ and assume that $h_2=\bar{h}_2=1$.  We look at terms in the OPE proportional to the scalar operator
 \begin{eqnarray}
 j^{n,n-1}(w,\bar{w}){\cal O}^{1,1}(0)= \frac{{\cal O}^{1,1}(0)}{w^{n}\bar{w}^{n-1}}+\cdots \nonumber\\
  j^{n-1,n}(w,\bar{w}){\cal O}^{1,1}(0)= \frac{{\cal O}^{1,1}(0)}{w^{n-1}\bar{w}^{n}}+\cdots 
 \end{eqnarray}
 When $n=1$ the smearing function (\ref{CSmap}) for a massless gauge field in AdS${}_{3}$ gives
 \begin{equation}
 A^{1,0}(z,w,\bar{w}) {\cal O}^{1,1}(0)=\frac{1}{w}{\cal O}^{1,1}(0) +\cdots
 \label{massless3point}
 \end{equation}
 On the other hand for a massive vector the smearing function (\ref{MassiveVectorSmear}) leads to
 \begin{equation}
 A^{n,n-1}(z,w,\bar{w}){\cal O}^{1,1}(0)=\left(-\frac{2}{\pi}\frac{d}{dw}I_{1}^{(n-1)}+\frac{z}{\pi}\frac{d}{dw}I_{2}^{(n)}\right){\cal O}^{1,1}(0)+\cdots
 \end{equation}
 where
 \begin{eqnarray}
 I_{1}^{(n-1)}&=&\int_{0}^{z}rdr\left( \frac{z^2-r^2}{z}\right)^{2n-3}\int_{|\alpha|=1}\frac{d\alpha}{\alpha(w+r\alpha)^{n-1}(\bar{w}-r/\alpha)^{n-1}}\nonumber\\
 I_{2}^{(n)}&=&\int_{0}^{z}rdr\left( \frac{z^2-r^2}{z}\right)^{2n-2}\int_{|\alpha|=1}\frac{d\alpha}{\alpha(w+r\alpha)^{n}(\bar{w}-r/\alpha)^{n}}
 \end{eqnarray}
 Using (\ref{iden1}) one gets
 \begin{eqnarray}
 I_{1}^{(n-1)}&=&\frac{\pi z^{2n-1}}{(2n-2)(w\bar{w})^{n-1}}F\Big(n-1,n-1,2n-1,-\frac{z^2}{w\bar{w}}\Big)\nonumber\\
 I_{1}^{(n-1)}&=&\frac{\pi z^{2n}}{(2n-1)(w\bar{w})^{n}}F\Big(n,n,2n,-\frac{z^2}{w\bar{w}}\Big)
 \end{eqnarray}
 and finally using (\ref{iden2}) one  gets
  \begin{equation}
 A^{n,n-1}(z,w,\bar{w}){\cal O}^{1,1}(0)=-{\cal O}^{1,1}(0)\frac{d}{dw}\left(\frac{z^{2n-1}}{(n-1)(w\bar{w})^{n-1}}F\big(n-1,n,2n-1,-\frac{z^2}{w\bar{w}}\big)\right)
 \label{3pointvector}
 \end{equation}
  A similar result holds for $A^{n-1,n}$ by replacing $w \rightarrow \bar{w}$.
  The quantity in parenthesis in (\ref{3pointvector}) is non-analytic due to terms of the form
 \begin{equation}
 \left(\frac{w\bar{w}}{z^2}\right)^{m} \ln \frac{z^2+w\bar{w}}{w\bar{w}}
 \label{3pnonanalytic}
 \end{equation}
  with $n \geq m\geq 1$.
  
  Suppose we have a massless gauge field in the bulk.  The singular term in (\ref{massless3point}) leads  to a non-vanishing commutator at bulk spacelike separation,
  and must be canceled if the gauge field is to commute at spacelike separation.  But given the structure (\ref{3pnonanalytic})
  there is no massive vector we can add to our definition of a bulk gauge field that will cancel the divergent term in
  (\ref{massless3point}). This means that it is not possible to promote a boundary conserved current to a local bulk field.\footnote{The lesson here is not that causality is violated.  For example in AdS${}_3$ the field strength associated with (\ref{CSmap}) vanishes identically, and in this
  sense micro-causality is trivially satisfied even in the presence of interactions.  Rather the lesson is that there is an obstacle to constructing bulk gauge fields
  which have local commutators.  This is a feature, not a bug, since as we discuss in section \ref{sect:gauge} gauge fields are expected to have non-local commutators.}
 
 On the other hand, starting from a non-conserved current in the CFT, there is no obstacle to restoring bulk locality. One can cancel non-analytic terms of the form
 (\ref{3pnonanalytic}) by adding a tower of higher-dimension spin-1 fields with appropriately chosen masses and coefficients to our definition of a bulk vector field.
 This will leave a non-analytic term of the form
 \begin{equation}
 \left(\frac{w\bar{w}}{z^2}\right)^{n_{\rm max}}\ln (w\bar{w})
 \end{equation}
 where $n_{\rm max}$ the largest $n$ used in the sum over higher dimension primaries. So, just as in the scalar case \cite{Kabat:2011rz}, we can make a massive vector field in
 the bulk as local as we wish.

\subsection{A comment on gauge fields\label{sect:gauge}}

If there is a gauge symmetry in the bulk, i.e.\ a conserved current on the boundary, the issue of constructing bulk operators become a bit more involved.  Of course one could start from the bulk equations of motion and solve them perturbatively,
to express bulk fields in terms of boundary data.  If one starts from a local bulk Lagrangian, this procedure is guaranteed to describe a local theory in the bulk
(at least perturbatively).  But if one wants to construct bulk operators purely in terms of the CFT, without making reference to bulk equations of motion, then having bulk gauge symmetries complicates matters.
If there is a gauge symmetry in the bulk then the corresponding charge can be expressed as a surface term and identified with a conserved
quantity in the CFT.  The charge generates global gauge transformations, so as discussed in \cite{Heemskerk:2012mn,Heemskerk:2012mq}, charged fields in the bulk must have non-local commutators in order to properly implement the Gauss constraint.  In the context of gravity this discussion
applies to time evolution, since the CFT Hamilton should generate time translation everywhere in the bulk.  While these non-local
commutators do not actually violate causality, they do
complicate the CFT construction, in the sense that the guiding principle of bulk causality must be stated more carefully.  It's tempting to speculate that the good causal properties we found for the field strength and Weyl tensor at the linearized
level can provide a basis for constructing the interacting theory, at least in perturbation theory.

\section{General backgrounds\label{sect:generalbackground}}

In a given fixed background one can solve the bulk equations of motion perturbatively, to write an
expression for the Heisenberg picture fields in the bulk in terms of the boundary
values of those same fields, now interpreted as operators in the dual CFT.
Correlation function of these CFT operators
then reproduce bulk correlation functions. The
computations are done from the bulk point of view in a particular
gauge $G_{z\mu}=0$, $G_{zz}=R^2/z^2$. With
gauge fields one also sets $A_{z}=0$. These conditions completely fix the
gauge.  The resulting computations are thus physical since
all redundant degrees of freedom have been eliminated. In a fixed
gauge one can reproduce bulk calculations using boundary data, and
since the boundary data comes from a unitary field theory this
constitutes holography. From the CFT point of view one corrects the
naive smeared operator (constructed to represent a free field in the bulk) by adding higher dimension smeared operators to get
a local bulk operator. However these calculations as presented are
done in a fixed background metric with a fixed causal structure. This
causal structure cannot be circumvented or changed in perturbation
theory since it is built in to the hardware of the
approach. The approach based on micro-causality and CFT correlators has the same
difficulty. One must define a smearing function which is determined by
the background metric, and this smearing function cannot be changed
in perturbation theory, aside from corrections to incorporate anomalous dimensions.

Besides the question: how local can bulk operators be in this
formalism?, one can ask how this formalism could work without an a
priori notion of a background. Here we make a few comments on
these issues.

In a fixed background the equations of motion for the bulk fields
come from a radial
Hamiltonian $H_{r}$.  (By radial Hamiltonian we mean the operator which generates radial evolution of fluctuations about this particular background.)  Schematically ($\phi$
stands for any perturbative field including gravitons on this
background)
\begin{equation}
\label{radial}
\frac{\partial \phi}{\partial z}=-[H_{r},\phi]
\end{equation}
We also need to impose an initial condition, given by normalizable falloff as $z
\rightarrow 0$ for each field.  The radial Hamiltonian can be
explicitly written down in the supergravity approximation. If we had a
different background metric then the radial Hamiltonian would be some
different operator, but for each background we can think of the radial
Hamiltonian as some operator in the CFT, generating the transformation
from boundary operators to bulk operators via the map
\begin{equation}
{\cal O}(x,t) \rightarrow e^{-\int_{0}^{z} H_{r}}  {\cal O}(x,t) e^{\int_{0}^{z} H_{r}}
\label{defbulkopr}
\end{equation}
However the idea that we will just get a different smearing function for each
background is still problematic. The construction of smearing functions
relies on having a classical spacetime (perhaps with a few perturbative
quantum fluctuations).  This clearly does not have to be the case for a
generic state in the CFT.

The approximation of getting a fixed background with a few supergravity
excitations on it involves two steps. First one needs to integrate out
all the bulk stringy modes, which in the CFT means integrating out all
high dimension operators. Second one must do a semiclassical approximation to get a well-defined background metric. We won't have much to say about the first
step, other than that one has to be careful later on when discussing high
dimension operators. For instance, in the promotion of a boundary
operator to a field in the bulk, one needs to include from the CFT perspective a
tower of high dimension operators. If one includes high dimension
operators only up to some $\Delta_{\rm max}$ then, according to \cite{Kabat:2011rz},
a good estimate of the commutator of a bulk operator with a boundary
operator (taken to be scalars in AdS$_{3}$), which are spacelike separated in the bulk but
not on the boundary, is
\begin{equation}
[\phi(t,{\bf x},z), {\cal O}(0)] \sim \Big(\frac{t^2-\vert {\bf x} \vert^2}{z^2}\Big)^{\Delta_{\rm max}}
\end{equation}
Although non-zero, the commutator is exponentially suppressed away from the bulk lightcone provided $\Delta_{\rm max}$ is large.
A nice way to characterize the bulk non-locality associated with a finite value of $\Delta_{\rm max}$ is to ask how far from the bulk lightcone one can go before the commutator
becomes exponentially small.  This is given by
\begin{equation}
\delta S \sim {R}/{\Delta_{\rm max}}
\end{equation}
where $R$ is the AdS radius and $S$ is proper length in the bulk. For
$\Delta_{\rm max} \sim (g_{\rm YM}^{2} N)^{1/4}$ -- appropriate for stringy modes --
one gets $\delta S \sim l_{s}$.

Even if the approximation of integrating out the stringy modes is good
it does not mean the CFT state describes a semiclassical space
time. In the supergravity approximation we can write down the
equations of motion for the metric and matter fields in holographic gauge
without choosing a particular background.
This is done by replacing the radial Hamiltonian in (\ref{radial}) with the appropriate
Hamiltonian for the supergravity system, namely $H_{g}=\int d^{d}x
\frac{1}{z^2}H_{WD}$ where $H_{WD}$ is the Wheeler -- de Witt
operator. The radial evolution equations are then
\begin{equation}
\label{WdW}
\frac{\partial {\cal O}}{\partial z}=-[H_{g},{\cal O}] \  \  \  \frac{\partial g_{\mu \nu}}{\partial z}=-[H_{g}, g_{\mu \nu}]
\end{equation}
and similarly for the conjugate momenta.  Once the constraints are
satisfied on the initial slice ($z=0$) the equations of motion
guarantee that they are obeyed at any $z$. We assume here that
\begin{equation}
g_{\mu \nu}(z \rightarrow 0)=\eta_{\mu \nu}
\end{equation}
So corrections to the bulk metric come from normalizable modes, with the
leading correction for small $z$ being proportional to $T_{\mu \nu}$. This
together with $\partial^{\mu} T_{\mu \nu}=0$ and $T^{\mu}_{\mu}=0$
gives enough initial data to solve the equations.\footnote{We are ignoring the question of whether holographic
gauge can be extended all the way to $z = \infty$.  Also since we are working in a Poincar\'e patch we are ignoring
any anomalous trace of the stress tensor.}

The equations of motion can formally be solved to give the bulk fields as functionals of
the boundary data.
\bea
&& \phi(x,z) = \phi(x,z)\big[T_{\mu\nu}(x'),{\cal O}(x'')\big] \\
\nonumber
&& g_{\mu\nu}(x,z) = g_{\mu\nu}(x,z)\big[T_{\mu\nu}(x'),{\cal O}(x'')\big]
\eea
So far this is independent of the state of the CFT. But now, given some state
of the CFT, we would like to obtain a set of bulk operators which look like
fields propagating on some semiclassical space time.  To do this, to a good approximation one needs to be able
to substitute
\begin{equation}
T_{\mu\nu}=\langle T_{\mu\nu}\rangle + \delta T_{\mu\nu}.
\end{equation}
If this approximation is valid then we are guaranteed that correlators of our bulk operators, calculated in the CFT,
will look like correlation
function of supergravity fields on a background which solves the
Einstein equations with asymptotics set by $\langle T_{\mu \nu} \rangle$.

Clearly such an approximation is valid in a CFT state if connected
correlation functions of CFT operators obey large $N$ factorization.
Thus CFT states with large $N$ factorization will be dual to
semiclassical spacetimes, while those which do not obey large $N$
factorization will not have a local spacetime interpretation.

Finally we want to speculate about a method for constructing bulk operators purely within the CFT. It seems possible from the above considerations that one can define a
master set of ``bulk operators'' in the CFT, regardless of the state of the CFT or any low energy approximation. These operators would not have a bulk interpretation, except on a restricted set of
states where 
large $N$ factorization holds. What are these master bulk operators? We propose to extrapolate from the supergravity situation (\ref{WdW}). A natural guess
is that they are defined by replacing the radial Hamiltonian in (\ref{defbulkopr}) with a more fundamental operator in the CFT, such as the exact RG Hamiltonian or Fokker-Planck Hamiltonian (see for instance \cite{Lifschytz:2000bj,Heemskerk:2010hk}).

\section{Conclusions\label{sect:conclusions}}

In this paper we worked out the smearing functions which describe linearized spin-1 and spin-2 excitations in AdS.  We showed that bulk locality is respected:
although gauge fields and metric perturbations have non-local commutators when one works in holographic gauge, the corresponding curvatures -- the field strength for $A_\mu$,
or the Weyl tensor in the case of gravity -- are causal.  We also studied massive vector fields, where the vector field itself is causal due to the non-conserved nature
of the dual boundary current.

These results could be extended in several directions.  For example we computed the smearing function for a Chern-Simons gauge field in AdS${}_3$.
It would be interesting to work out the smearing function for a Maxwell field in AdS${}_3$, dual to a CFT with a dynamical gauge field \cite{Marolf:2006nd,Jensen:2010em}
(see however \cite{Andrade:2011dg}).
Our results could be used to study the Maxwell-Chern-Simons theory recently analyzed in \cite{Andrade:2011sx}.  Since the smearing functions are basically fixed by
AdS covariance, our results should also apply if there is a duality between AdS${}_2$ and CFT${}_1$, although the physical interpretation in this context is not so clear.

Perhaps a more interesting direction is to extend our results to include interactions.  For massive vector fields we showed how this works in section \ref{sect:interaction}:
in a $1/N$ expansion one adds appropriately smeared higher-dimension vector operators, with coefficients that are fixed by the requirement of bulk causality.  It would be
very interesting to extend this to gauge fields and metric perturbations, perhaps using the good causal properties of the field strength and Weyl tensor as a guiding principle.
Ultimately one might hope to make contact between the `bottom-up' approach of constructing bulk observables in $1/N$ perturbation theory, and the `top-down' approach
of section \ref{sect:generalbackground} where bulk operators are constructed from a fundamental operator of the boundary theory.

\bigskip
\goodbreak
\centerline{\bf Acknowledgements}
\noindent
We are grateful to Idse Heemskerk, Don Marolf, Joe Polchinski and James Sully for valuable discussions.  The work of DK, SR and DS was supported by
U.S.\ National Science Foundation grant PHY-0855582 and by PSC-CUNY grants. The work of GL was supported in part by the Israel Science Foundation under Grant No.\ 392/09.
This research was  supported in part by the National Science Foundation under Grant No.\ PHY11-25915.

\appendix
\section{Scalar smearing functions\label{sect:scalar}}

Consider a scalar field of mass $m$ in AdS${}_{d+1}$.  It's dual to an operator of dimension $\Delta$ in the CFT,
where $m^2 R^2 = \Delta (\Delta - d)$.  The mode expansion is
\be
\label{ScalarMode}
\phi(t,{\bf x},z) = \int_{\vert \omega \vert > \vert {\bf k} \vert} d\omega d^{d-1}k \, a_{\omega {\bf k}} e^{-i \omega t} e^{i {\bf k} \cdot {\bf x}}
z^{d/2} J_\nu(z\sqrt{\omega^2 - \vert {\bf k} \vert^2})
\ee
where $\nu = \Delta - d/2$.  As $z \rightarrow 0$ we have $\phi(t,{\bf x},z) \sim z^\Delta \phi_0(t,{\bf x})$ where the boundary field
\[
\phi_0(t,{\bf x}) = {1 \over 2^\nu \Gamma(\nu + 1)} \int_{\vert \omega \vert > \vert {\bf k} \vert} d\omega d^{d-1}k \,
a_{\omega {\bf k}} e^{-i \omega t} e^{i {\bf k} \cdot {\bf x}} (\omega^2 - \vert {\bf k} \vert^2)^{\nu/2}
\]
Our basic goal is to express the bulk field in terms of the boundary field.  A straightforward way to do this is to express the coefficients
$a_{\omega {\bf k}}$ as a Fourier transform of $\phi_0$,
\[
a_{\omega {\bf k}} = {2^\nu \Gamma(\nu + 1) \over (2\pi)^d (\omega^2 - \vert {\bf k} \vert^2)^{\nu/2}} \int dt d^{d-1} x \,
e^{i \omega t} e^{-i {\bf k} \cdot {\bf x}} \phi_0(t,{\bf x})\,.
\]
Substituting this back in (\ref{ScalarMode}) leads to an integral representation of the smearing function.  Generically one obtains a smearing
function with support on the entire boundary of the Poincar\'e patch, however by complexifying the boundary spatial coordinates one can obtain
a smearing function with compact support.  As shown in \cite{Hamilton:2006fh} this leads to
\be
\label{ScalarSmear}
\phi(t,{\bf x},z) = {\Gamma(\Delta - {d \over 2} + 1) \over \pi^{d/2} \Gamma(\Delta - d + 1)}
\hspace{-7mm} \int\limits_{\hspace{8mm}t'{}^2 + \vert {\bf y}'\vert^2 < z^2} \hspace{-5mm} dt' d^{d-1} y' \,
\left({z^2 - t'{}^2 - \vert {\bf y'} \vert^2 \over z}\right)^{\Delta - d} \phi_0(t + t', {\bf x} + i {\bf y}')
\ee
This expression is fine for $\Delta > d-1$.  However when $\Delta = d-1$ it's ill-defined: the integral diverges, and the
coefficient in front goes to zero.

To construct a smearing function for $\Delta = d - 1$ we return to the mode expansion (\ref{ScalarMode}).
As a warm-up example take a massless field in AdS${}_2$ with $\Delta = 0$.  The mode expansion is
$\phi(t,z) = \int d\omega \, a_\omega e^{-i \omega t} \cos(\omega z)$.
Then
$
a_\omega = {1 \over 2 \pi} \int dt \, e^{i \omega t} \phi_0(t)
$
and
\bea
\nonumber
\phi(t,z) & = & \int dt' \int {d\omega \over 2 \pi} e^{-i \omega (t - t')} \cos(\omega z) \phi_0(t') \\
& = & {1 \over 2} \left(\phi_0(t+z) + \phi_0(t-z)\right)
\eea
This clearly satisfies the wave equation $(\partial_t^2 - \partial_x^2)\phi = 0$ and obeys
the boundary condition $\phi(t,z) \rightarrow \phi_0(t)$
as $z \rightarrow 0$.  It can be written in the covariant form
\[
\phi(t,z) = {1 \over 2} \int dt' \, \delta(\sigma z') \phi_0(t')
\]
where $\sigma z' = {z^2 - (t - t')^2 \over 2z}$.

We now consider the general case of a field with $\Delta = d - 1$.  In any dimension solving for $a_{\omega {\bf k}}$ in terms of $\phi_0$
and plugging back into the mode expansion gives
\be
\label{anydim}
\phi(t,{\bf x},z) = \int_{\vert \omega \vert > \vert {\bf k} \vert} d\omega d^{d-1}k \,
{2^\nu \Gamma(d/2) z^{d/2} \over (2\pi)^d (\omega^2 - \vert {\bf k} \vert^2)^{\nu/2}} \,
J_\nu\big(z\sqrt{\omega^2 - \vert {\bf k} \vert^2}\big) e^{-i \omega t} e^{i {\bf k} \cdot {\bf x}} \phi_0(\omega,{\bf k})
\ee
Here $\nu = {d \over 2} - 1$ and $\phi_0(\omega,{\bf k})$ is the Fourier transform of the boundary field.  The Bessel function
has an integral representation
\be
J_\nu(a) = {1 \over \sqrt{\pi} \, \Gamma\left(\nu + {1 \over 2}\right)} \left({a \over 2}\right)^\nu \int_0^\pi d\theta \,
e^{-i a \cos \theta} \sin^{2\nu} \theta
\ee
or equivalently
\be
J_\nu(a) = {1 \over \sqrt{\pi} \, \Gamma\left(\nu + {1 \over 2}\right)} \left({a \over 2}\right)^\nu \, {1 \over {\rm vol}(S^{d-2})}
\int_{\vert {\bf n} \vert = 1} d{\bf n} \, e^{-i {\bf a} \cdot {\bf n}}
\ee
Here ${\bf a}$ is a $d$-component vector with Euclidean norm $a$ and ${\bf n} \in S^{d-1}$ is a unit vector.
Setting ${\bf a} = z (\omega,-ik_1,\ldots,-ik_{d-1})$ and using
\be
{\rm vol}(S^{d-1}) = {2 \pi^{d/2} \over \Gamma(d/2)}
= {\sqrt{\pi} \, \Gamma\left({d-1 \over 2}\right) {\rm vol}(S^{d-2}) \over \Gamma(d/2)}
\ee
this becomes
\[
{2^\nu \Gamma(d/2) z^{d/2} \over (\omega^2 - \vert {\bf k} \vert^2)^{\nu/2}} \, J_\nu(z\sqrt{\omega^2 - \vert {\bf k} \vert^2}) =
{1 \over {\rm vol}(S^{d-1})} \hspace{-7mm} \int\limits_{\hspace{8mm}t'{}^2 + \vert {\bf y}'\vert^2 = z^2} \hspace{-5mm} dt' d^{d-1} y' \,
e^{-i \omega t'} e^{-{\bf k} \cdot {\bf y}'}
\]
Using this representation in (\ref{anydim}) leads to\footnote{The boundary field $\phi_0$ only has Fourier components with $\vert \omega \vert > \vert {\bf k} \vert$, so we can integrate over $\omega$ and ${\bf k}$ without restriction.}
\be
\label{continue}
\phi(t,{\bf x},z) = {1 \over {\rm vol}(S^{d-1})} \hspace{-7mm} \int\limits_{\hspace{8mm}t'{}^2 + \vert {\bf y}'\vert^2 = z^2} \hspace{-5mm} dt' d^{d-1} y' \,
\int {d\omega d^{d-1} k \over (2 \pi)^d} e^{-i \omega (t + t')} e^{i {\bf k} \cdot ({\bf x} + i {\bf y}')} \phi_0(\omega,{\bf k})
\ee
We interpret the Fourier transforms in (\ref{continue}) as defining the analytic continuation of $\phi_0(t,{\bf x})$ to complex ${\bf x}$.  Thus the smearing function for a
scalar field with $\Delta = d - 1$ is
\be
\label{scalarsmear}
\phi(t,{\bf x},z) = {1 \over {\rm vol}(S^{d-1})} \hspace{-7mm}
\int\limits_{\hspace{8mm}t'{}^2 + \vert {\bf y}'\vert^2 = z^2} \hspace{-5mm} dt' d^{d-1} y' \,
\phi_0(t + t', {\bf x} + i {\bf y}')
\ee
This can be written in a covariant form
\be
\label{scalarsmear2}
\phi(t,{\bf x},z) = {1 \over {\rm vol}(S^{d-1})} \int dt' d^{d-1} y' \, \delta(\sigma z') \, \phi_0(t + t', {\bf x} + i {\bf y}')
\ee
in terms of the bulk - boundary distance (\ref{sigmaz'}).

It's clear that (\ref{scalarsmear}), (\ref{scalarsmear2}) satisfy the correct boundary conditions.  As $z \rightarrow 0$ the integration
region on the boundary shrinks to a point, so we can bring the boundary field outside the integral and recover
\[
\phi(t,{\bf x},z) \sim z^{d-1} \phi_0(t,{\bf x}) \qquad \hbox{\rm as $z \rightarrow 0$}
\]
One can also check that (\ref{scalarsmear2}) satisfies the wave equation.  Acting on a function of the AdS-invariant distance $\sigma$,
the wave equation $(\Box - m^2) \phi = 0$ reduces to
\[
(\sigma^2 - 1) \phi'' + (d + 1) \sigma \phi' - \Delta(\Delta - d) \phi = 0
\]
With a small fixed cutoff $z'$, the smearing kernel appearing in (\ref{scalarsmear2}) is ${1 \over z'} \delta(\sigma)$.  We want to check
that this is annihilated by the wave operator in the limit $z' \rightarrow 0$.  To do this we act with the wave operator and integrate against
a test function $f(\sigma z')$ (the test function can be thought of as the boundary field).  For $\Delta = d - 1$ this gives
\beas
& & \int d(\sigma z') \, f(\sigma z') \left[(\sigma^2 - 1) {d^2 \over d\sigma^2} + (d+1) \sigma {d \over d\sigma} + (d-1)\right] {1 \over z'} \delta(\sigma) \\
&=& \int d(\sigma z') {1 \over z'} \delta(\sigma) \left[{d^2 \over d\sigma^2} (\sigma^2 - 1) - (d+1) {d \over d\sigma} \sigma + (d-1) \right] f(\sigma z') \\
&=& - z'{}^2 f''(0)
\eeas
This vanishes as $z' \rightarrow 0$, which shows that the wave equation is satisfied when the regulator is removed.

\section{Chern-Simons in holographic gauge\label{appendix:CS}}

Our goal in this appendix is to quantize Chern-Simons theory in holographic gauge.  We want to show
that we recover the bulk commutator (\ref{CSbulk}) obtained in section \ref{sect:CS} by applying our smearing
functions to the current algebra on the boundary.

We begin from the abelian Chern-Simons action\footnote{Conventions: light-front coordinates
are $x^\pm = t \pm x$.  We take $\epsilon_{012} = +1$ and relate the bulk and boundary
orientations by $\int d^3x \, \partial_z f = - \int d^2x f \vert_{z = 0}$.}
\[
S_{\rm bulk} = \int d^3x \, {1 \over 2} \kappa \, \epsilon^{ABC} A_A \partial_B A_C
\]
To obtain a right-moving current algebra on the boundary we supplement this with a surface
term \cite{Kraus:2006wn}
\[
S_{\rm bdy} = \int d^2x \, \kappa A_+ A_-
\]
The surface term leads to a well-defined variational principle provided we impose the boundary
condition that $A_-$ is fixed (that is, $\delta A_- = 0$) on the boundary.

In light-front coordinates one can integrate by parts to find (the surface terms cancel against $S_{\rm bdy}$)
\[
S_{\rm bulk + bdy} = \int d\xp d\xm dz \, \kappa A_z \partial_+ A_-
+ \kappa A_+ \left(\partial_- A_z - \partial_z A_-\right)\,.
\]
We adopt $\xp$ as light-front time \cite{Heinzl:2000ht} and read off the
Poisson bracket \cite{Faddeev:1988qp}
\[
\lbrace A_z(\xm,z), A_-(\xmp,z') \rbrace = {1 \over \kappa} \delta(\xm - \xmp) \delta(z - z')
\]
$A_+$ is a Lagrange multiplier that enforces the Chern-Simons Gauss law.  Thus we
have a (primary, first-class) constraint
\[
\chi_1 = \partial_z A_- - \partial_- A_z \approx 0\,.
\]
The constraint generates the expected gauge transformation
\beas
&& \delta A_z = \left\lbrace \int d\xmp dz' \, \lambda_1 \chi_1,\, A_z(\xm,z) \right\rbrace
= {1 \over \kappa} \partial_z \lambda_1 \\
&& \delta A_- = \left\lbrace \int d\xmp dz' \, \lambda_1 \chi_1,\, A_-(\xm,z) \right\rbrace
= {1 \over \kappa} \partial_- \lambda_1
\eeas
To preserve the boundary condition $\delta A_- \vert_{z = 0} = 0$, we require that the gauge parameter
satisfy $\lambda_1 \vert_{z = 0} = 0$.  We wish to work in holographic gauge, so we impose an additional
constraint (a gauge-fixing condition)
\[
\chi_2 = A_z \approx 0\,.
\]
The constraints obey
\[
\Delta_{ij} \equiv \lbrace \chi_i,\, \chi_j \rbrace = \left(
\begin{array}{cc}
0 & - {1 \over \kappa} \delta(\xm - \xmp) \delta'(z - z') \\
- {1 \over \kappa} \delta(\xm - \xmp) \delta'(z - z') & 0
\end{array}\right)
\]
Acting on functions $\left({\lambda_1 \atop \lambda_2}\right)$ this operator has zero
modes, but as we will see the zero modes can be eliminated by requiring
\[
\lambda_1(x^-,z = 0) = 0 \qquad\quad \lambda_2(x^-,z = \infty) = 0
\]
Then $\Delta$ has a well-defined inverse,
\[
\Delta^{-1} = \left(
\begin{array}{cc}
0 & - \kappa \delta(\xm - \xmp) \theta(z - z') \\
\kappa \delta(\xm - \xmp) \theta(z' - z) & 0
\end{array}\right)
\]
Note that $\Delta^{-1}$ is antisymmetric.  One can easily check the basic property
\[
\Delta^{-1} \Delta \left({\lambda_1 \atop \lambda_2}\right) =
\left(\begin{array}{l} \lambda_1(\xm,z) - \lambda_1(\xm,0) \\
\lambda_2(\xm,z) - \lambda_2(\xm,\infty) \end{array}\right)
\]
which shows that $\Delta$ is invertible given our boundary conditions.
The constraints can be eliminated by defining Dirac brackets.  The Dirac bracket of $A_z$
with anything will vanish, while the Dirac bracket of $A_-$ with itself is
\beas
\left\lbrace A_-(\xm,z),\, A_-(\xmp,z') \right\rbrace & = &
0 - \left\lbrace A_-,\, \chi_i \right\rbrace \Delta^{-1}_{ij} \left\lbrace \chi_j,\, A_- \right\rbrace \\
& = & - {1 \over \kappa} \delta'(\xm - \xmp)
\eeas
Quantizing via $\lbrace \cdot,\,\cdot \rbrace \rightarrow i [ \cdot,\,\cdot ]$ reproduces the bulk commutator (\ref{CSbulk})
and fixes the normalization $\kappa = 4 \pi / k$.


\providecommand{\href}[2]{#2}\begingroup\raggedright\endgroup

\end{document}